\documentclass[aps,prd,twocolumn,superscriptaddress,longbibliography,nofootinbib]{revtex4-2}
\usepackage[utf8]{inputenc}
\usepackage{color}
\usepackage{graphicx}
\usepackage{subfigure}
\usepackage{xcolor}
\usepackage[normalem]{ulem}
\usepackage[pdftex,breaklinks,colorlinks,linkcolor=blue,citecolor=teal,anchorcolor=red,urlcolor=cyan]{hyperref}
\usepackage{orcidlink}
\usepackage{multirow}

\usepackage{amsmath,amssymb,amsfonts}

\def\prl{Phys. Rev. Lett.}
\def\prd{Phys. Rev. D}

\def\pau_p{Prog. Theor. Phys.}

\def\I{\mathcal I}
\def\J{\mathcal J}

\begin{document}
\title{Critical collapse of vacuum spacetimes: Nakamura wave initial data}

\author{Thomas W. Baumgarte\orcidlink{0000-0002-6316-602X}}
\email{tbaumgar@bowdoin.edu}
\affiliation{Department of Physics and Astronomy, Bowdoin College, Brunswick, Maine 04011, USA}

\author{Carsten Gundlach\orcidlink{0000-0001-9585-5375}}
\email{c.j.gundlach@soton.ac.uk}
\affiliation{School of Mathematical Sciences, University of Southampton,
  Southampton SO17 1BJ, United Kingdom} 

\author{David Hilditch\orcidlink{0000-0001-9960-5293}}
\email{david.hilditch@tecnico.ulisboa.pt}
\affiliation{
  Centro de Astrof\'{\i}sica e Gravita\c c\~ao -- CENTRA,
  Departamento de F\'{\i}sica, Instituto Superior T\'ecnico -- IST,
  Universidade de Lisboa -- UL, Av.\ Rovisco Pais 1, 1049-001 Lisboa,
  Portugal}

\begin{abstract}
We report on numerical simulations of critical phenomena in the collapse of axisymmetric vacuum gravitational waves, adopting families of initial data that, to the best of our knowledge, have not been used in this context before.  Like Teukolsky waves, the data are based on linear wave solutions to the Einstein equations.  We follow Nakamura's construction and encode the wave content in the extrinsic curvature rather than the spatial curvature, which leads to several simplifications when the data are ``dressed up" so that they satisfy the nonlinear constraint equations.  We are able to fine-tune these data to the onset of black hole formation slightly better than in our previous simulations, allowing us to observe and examine one more echo in the approximately self-similar threshold solution.  Our findings are consistent with earlier studies: while we find threshold solutions that are approximately discretely self-similar, the self-similarity is not exact, and we find no evidence for a unique critical solution.  We discuss common features between the different threshold solutions, including the appearance of alternating maxima in the direction of the poles and the equator.
\end{abstract}
\maketitle

%

%
\section{Introduction}
\label{sec:intro}
%

Critical phenomena in gravitational collapse -- first reported in the seminal work by Choptuik \cite{Cho93} -- refer to properties of solutions to Einstein's equations in the vicinity of the threshold of black hole formation.  Specifically, consider a family of initial data parametrized by a parameter $\eta$, say.  A critical value of this parameter, $\eta_*$, separates {\em supercritical} data, which evolve to form black holes, from {\em subcritical} data that do not.  For supercritical data with $\eta$ sufficiently close to $\eta_*$, the mass of the forming black holes displays power-law {\em scaling},
\begin{equation} \label{eq:scaling_law}
M \propto |\eta - \eta_*|^{\gamma},
\end{equation}
where $\gamma$ is a {\em critical exponent} that is {\em universal} for a given matter model.  For subcritical data, other dimensional measures characterizing the spacetime evolution, e.g.~the maximum attained spacetime curvature, follow similar scaling laws (see \cite{GarD98}).  

Fine-tuning the data to the critical parameter $\eta_*$ results in a critical solution that {\em self-similarly} contracts towards an accumulation event.  The critical solution is again universal for a given matter model.  As demonstrated by \cite{KoiHA95,Mai96}, the critical exponent $\gamma$ is the inverse of the Lyapunov exponent that governs the single unstable mode of the critical solution -- the universality of $\gamma$ is therefore directly related to that of the critical solution. 

Depending on the matter model, the critical solution can be {\em continuously} self-similar (CSS), e.g.~for ideal fluids (see \cite{EvaC94}), or {\em discretely} self-similar (DSS), as for the scalar fields originally considered by Choptuik \cite{Cho93}.   A DSS solution can be pictured as having an oscillation superimposed on the self-similar contraction.  The period of this oscillation is referred to as the {\em echoing period}, and we note that, for a DSS critical solution, the scaling law (\ref{eq:scaling_law}) will also have a periodic fine-structure superimposed on the power-law (see \cite{HodP97,Gun97}).  

Choptuik's original simulations \cite{Cho93}, as well as numerous follow-up calculations (see, e.g., \cite{GunHMG25} for a review), were performed in spherical symmetry, raising the question whether the above picture of critical collapse also generalizes to the absence of spherical symmetry.  Remarkably, Abrahams and Evans \cite{AbrE93,AbrE94} reported the first results on critical collapse in axisymmetry only months after Choptuik's original announcement.  Specifically, they considered the collapse of vacuum gravitational waves and found an at least approximately DSS critical solution as well as some evidence for both universality and scaling.  In fact, the value for $\gamma$ reported by Abrahams and Evans \cite{AbrE93,AbrE94} is very similar to that reported by Choptuik \cite{Cho93} for scalar fields.  Given these observations, it became generally expected that the properties of critical collapse in spherical symmetry -- in particular the existence of a unique self-similar critical solution with an associated unique critical exponent -- should continue to hold even in the absence of spherical symmetry.

As we will discuss in more detail below, it has been difficult to reproduce the results of \cite{AbrE93,AbrE94}.  In the meantime, a series of other calculations in the absence of spherical symmetry also paint a more complicated picture.  Choptuik {\it et.al.}~\cite{ChoHLP03} considered a scalar field in axisymmetry and found that, for large deformations from spherical symmetry and for exquisite fine-tuning, the critical solution develops curvature maxima away from the center of symmetry, suggesting a possible bifurcation (compare \cite{Bau18,MarCRAH24} but also \cite{GunBH24}).  In another example, we considered the collapse of dipolar electromagnetic waves and found an approximately but not exactly DSS critical solution (see \cite{BauGH19}).  Generalizing these results for quadrupolar waves appears to result in a distinct threshold solution with curvature maxima away from the center, akin to those found for the scalar fields discussed above (see \cite{PerB21}).  We will refer to the limiting solution as the ``threshold solutions", rather than ``critical solution", if it is not unique.  Corresponding to the different threshold solutions, the dipolar and quadrupolar families also appear to feature approximate scaling with different exponents $\gamma$ (but see also \cite{ReiC23}, who report that the dipolar exponent will dominate for better fine-tuning even for quadrupolar initial data).  Taken together, there currently is no evidence that critical collapse in the absence of spherical symmetry leads to a unique and exactly self-similar critical solution (with corresponding unique critical exponent). 

Attempts to reproduce the results of Abrahams and Evans \cite{AbrE93,AbrE94} (see, e.g., \cite{Alcetal00,GarD01,Rin08,Sor11,Hiletal13}) have been hampered by a number of different challenges.  One problem has been that many modern numerical relativity simulations adopt 1+log slicing (see \cite{BonMSS95}).  As pointed out in \cite{Alc97,AlcM98,Alc05}, this slicing condition can develop coordinate shocks.  While this does not appear to happen in simulations of black holes and neutron stars, for which 1+log slicing has been extremely successful, such shocks do form in simulations of vacuum gravitational waves (see, e.g., \cite{Hiletal13,BauGH23}).  Rather than using 1+log slicing, recent successful simulations of vacuum critical collapse have employed a quasi-maximal slicing condition (\cite{KhiL18}, see \cite{LedK21}), a choice of gauge-source functions in the context of the generalized harmonic formulation (see \cite{FerRABH22}), or the shock-avoiding slicing condition suggested by Alcubierre (\cite{Alc03}, see \cite{BauGH23}).  Simulations with these three slicing conditions have been compared directly in \cite{Bauetal23}.

Another challenge is related to the choice of initial data.  For their work, Abrahams and Evans \cite{AbrE93,AbrE94} adopted Teukolsky-like initial data, i.e.~data that are based on the exact linear wave solutions of Teukolsky \cite{Teu82} but are then ``dressed up" to satisfy the non-linear constraint equations (see Section \ref{sec:teukolsky} below).  The latter step typically involves inverting curved and non-linear Laplace operators, which can be complicated (see also the discussion in \cite{LedK21,Ros25}).  Many authors therefore adopted Brill wave initial data \cite{Bri59} instead, which entails solving a flat and linear elliptic equation only and hence is considerably easier, and which, assuming universality, should result in the same results.  However, it now appears that there is no unique critical solution for the collapse of gravitational waves.  The authors of \cite{LedK21} found different critical exponents for different families of Teukolsky and Brill wave initial data, those of \cite{FerRABH22} compared different families of Brill wave data and also found different critical exponents, and in \cite{BauGH23} we compared quadrupolar and hexadecapolar Teukolsky waves and again found distinct threshold solution.   Moreover, many authors had considered geometrically prolate Brill waves (with a positive-amplitude seed function, see the discussion in \cite{Hiletal13}).  Among those families studied to date, these data appear to be the most complicated to analyze (see, e.g., \cite{LedK21,FerRABH22,Bauetal23}), further complicating a comparison with the results of Abrahams and Evans \cite{AbrE93,AbrE94}.

The picture that has emerged from these studies of vacuum critical collapse is consistent with that for non-vacuum critical collapse in axisymmetry discussed above: namely, at least at the current level of fine-tuning, there is no evidence for a unique critical solution (with a corresponding unique critical exponent), nor do the threshold solutions for any of the families studied so far display an exact self-similarity.

The purpose of this paper is to explore critical collapse of gravitational waves for a family of initial data that, to the best of our knowledge, has not yet been considered in this context.  Like the Teukolsky-like data discussed above, these data are based on analytical wavelike solutions to the linearized Einstein equations, but rather than using the analytical expressions to construct the spatial metric, they are used to construct the extrinsic curvature.   Since a similar approach was suggested by Nakamura \cite{Nak84} (see also \cite{ShiN95}), we will refer to this approach as ``Nakamura waves".  As we will discuss in Section \ref{sec:indata} below, this approach leads to several simplifications in the construction of the (nonlinear) initial data.

Our paper is organized as follows.  In Sect.~\ref{sec:indata} we review the construction of both Teukolsky and Nakamura wave initial data, we describe our numerical approach in Sect.~\ref{sec:numerics}, we present our numerical results in Sect.~\ref{sec:results}, and we briefly summarize in Sect.~\ref{sec:discussion}.  Throughout this paper we use geometrized units with $G = c = 1$.

%
\section{Gravitational wave initial data}
\label{sec:indata}
%

Numerical simulations of vacuum critical collapse require initial data describing nonlinear gravitational waves, i.e.~vacuum solutions to the Hamiltonian constraint 
\begin{equation} \label{eq:ham}
\bar D^2 \psi - \frac{\psi}{8} \bar R - \frac{1}{12} \psi^5 K^2 + \frac{1}{8} \psi^{-7} \hat A_{ij} \hat A^{ij} = 0
\end{equation}
and the momentum constraint
\begin{equation} \label{eq:mom}
\bar D_j \hat A^{ij} - \frac{2}{3} \psi^6 \bar \gamma^{ij} \bar D_j K = 0.
\end{equation}
Here $\bar \gamma_{ij} = \psi^{-4} \gamma_{ij}$ is the conformally related spatial metric (and $\gamma_{ij}$ the spatial metric induced by the spacetime metric on the initial slice), $\psi$ the conformal factor, and $\bar D_i$ and $\bar R$ the covariant derivative and Ricci scalar associated with $\bar \gamma_{ij}$.  We also decompose the extrinsic curvature
\begin{equation} \label{eq:extrinsiccurvature}
K_{ij} = A_{ij} + \frac{1}{3} \gamma_{ij} K = 
\psi^{-2} \hat A_{ij} + \frac{1}{3} \gamma_{ij} K
\end{equation}
into its trace-free part $A_{ij}$ and trace $K$, and conformally rescale the former according to $A_{ij} = \psi^{-2} \hat A_{ij}$.\footnote{We use a hat to distinguish this choice of conformal rescaling, which is commonly adopted in the context of the initial value problem, from the rescaling $A_{ij} = \psi^4 \bar A_{ij}$ used in the context of some evolution formulations.}  

Typically, nonlinear gravitational-wave solutions to the constraints (\ref{eq:ham}) and (\ref{eq:mom})  are constructed using one of two approaches, namely using Brill waves \cite{Bri59} -- which are constructed from a deformation of the conformally related metric -- or ``dressing up" a linear wave solution so that it satisfies the nonlinear constraint equations.  Here we focus on two different options for the latter approach, namely data based on the linear wave solutions of Teukolsky \cite{Teu82} and those of Nakamura \cite{Nak84}.  Both of these approaches construct linear perturbations of the flat Minkowski spacetime in transverse-traceless gauge, and both result in maximal slicing with $K = 0$.  In the following we will focus on even-parity solutions only.

%
\subsection{Teukolsky waves}
\label{sec:teukolsky}
%

Teukolsky waves (\cite{Teu82}, see also \cite{Rin09} for generalizations for all multipole moments) describe linearized waves propagating in a flat background metric.  The solutions are constructed from two seed functions $F_+(r+t)$ and $F_-(r-t)$ that can only depend on the combinations $r+t$ and $r-t$ and describe ingoing and outgoing waves.  From $F_+$ and $F_-$ three new functions $A$, $B$, and $C$ are computed, which, for quadrupolar waves ($\ell = 2$), entails up to four radial derivatives of the seed functions.  The functions $A$, $B$, and $C$ are then multiplied with certain angular functions (related to scalar, vector, and tensor spherical harmonics) to construct a spacetime metric that satisfies Einstein's equations to linear order in the wave amplitude.  From the spacetime metric we can identify the spatial metric $\gamma^{\rm Teu}_{ij}$ and the extrinsic curvature $\hat A_{ij}^{\rm Teu}$ (computed from the time derivative of $\gamma^{\rm Teu}_{ij}$).   Since the spacetime metric is derived from the linearized Einstein equations, $\gamma^{\rm Teu}_{ij}$ and $\hat A_{ij}^{\rm Teu}$ will generally satisfy the constraints (\ref{eq:ham}) and (\ref{eq:mom}) to linear order in the wave amplitude only.

Different authors have used different approaches to ``dress up" the linear solution so that it satisfies the constraint equations (\ref{eq:ham}) and (\ref{eq:mom}) nonlinearly, i.e.~to all orders in the wave amplitude (see \cite{AbrE93,AbrE94,LedK21,Ros25} for discussions).  In \cite{BauGH23} we adopted the following approach, following the convention and notation of \cite{Rin09} (see also Appendix A in \cite{BauGH23} for details).  We chose the seed function $F$ to be a linear combination of $F_+$ and $F_-$ so that the time derivative of the combination vanishes at $t = 0$, meaning that the initial slice represents a moment of time symmetry.  Specifically, we chose
\begin{equation} \label{eq:seed_teukolsky}
F(t,r) = \frac{{\mathcal A}}{2}
\left(u \left(e^{-u_+^2} + e^{-u_-^2}\right) 
+ v \left(e^{-v_+^2} + e^{-v_-^2}\right) \right),
\end{equation}
where ${\mathcal A}$ parametrizes the wave amplitude, and where $u = r-t$ and $v = r+t$, as well as $u_\pm = r - t \pm r_0$ and $v_\pm = r + t \pm r_0$, all in the dimensionless units adopted in our code.  The value of $r_0$ is related to the location of the peak of the resulting wave packet (allowing ``off-centered" initial data) and was chosen $r_0 = 2$ for all simulations presented in \cite{BauGH23}, and referred to again below.  We considered both quadrupolar ($\ell = 2$) and hexadecapolar ($\ell = 4$) solutions in \cite{BauGH23}, but will focus on the quadrupolar waves here.  We also adopted axisymmetry ($m = 0$), and note that the solutions satisfy a symmetry across the equatorial plane.  

Given that $\partial_t F(0,r) = 0$, this construction results in a moment of time symmetry at $t = 0$, meaning that the extrinsic curvature vanishes exactly at that moment, $\hat A_{ij}^{\rm Teu} = 0$. The momentum constraint (\ref{eq:mom}) is then satisfied identically, and only the Hamiltonian constraint (\ref{eq:ham}) remains to be solved.  Towards that end, we identified the conformally related metric with the wave metric, i.e.~$\bar \gamma_{ij} = \gamma^{\rm Teu}_{ij}$, then solved (\ref{eq:ham}) with this choice of $\bar \gamma_{ij}$ to find the conformal factor $\psi$, and finally computed the physical spatial metric $\gamma_{ij} = \psi^4 \bar \gamma_{ij}$.

Note, however, that this approach leads to several complications.  Since the spatial metric is not conformally flat, the term $\bar R$ in (\ref{eq:ham}) is non-zero, and, more importantly, the Laplace operator $\bar D^2$ is {\em not} a flat Laplace operator.  Another complication arises in the context of the so-called Baumgarte-Shapiro-Shibata-Nakamura (BSSN) formalism \cite{ShiN95,BauS98} which adopts the conformal connection functions $\bar \Gamma^i = \bar \gamma^{jk} \bar \Gamma^i_{jk}$ as a dynamical variable.  Computing initial data for $\bar \Gamma^i$ entails derivatives of the spatial metric, and hence additional derivatives of the seed function $F$.

\begin{figure*}
\centering
Teukolsky\hspace{1.5in}
Nakamura A\hspace{1.5in}
Nakamura B

\includegraphics[width = 0.31 \textwidth]{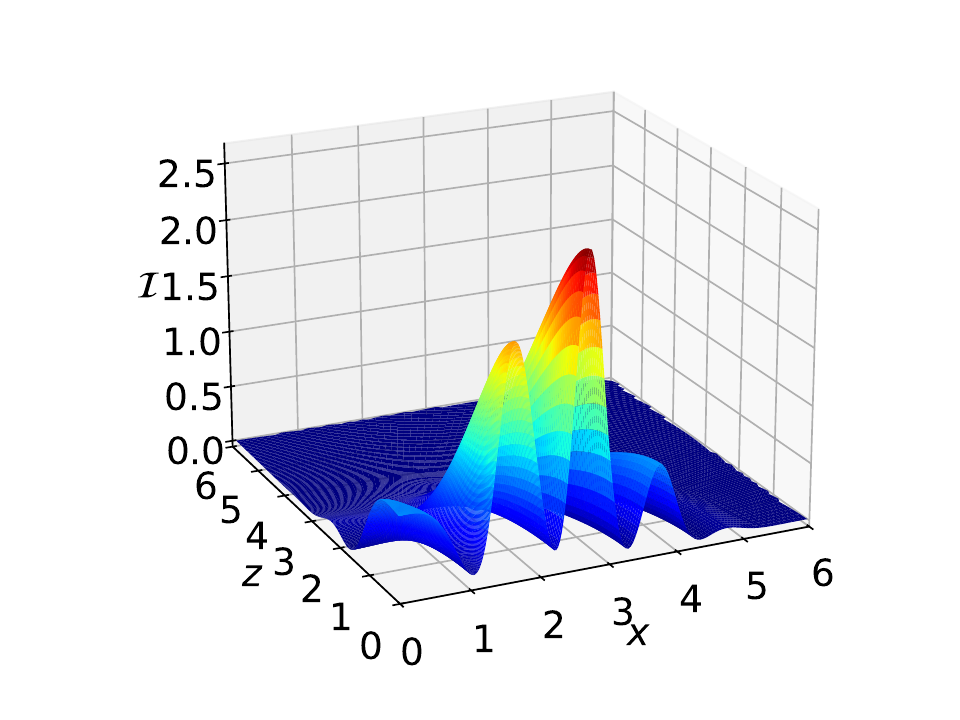}
\includegraphics[width = 0.31 \textwidth]{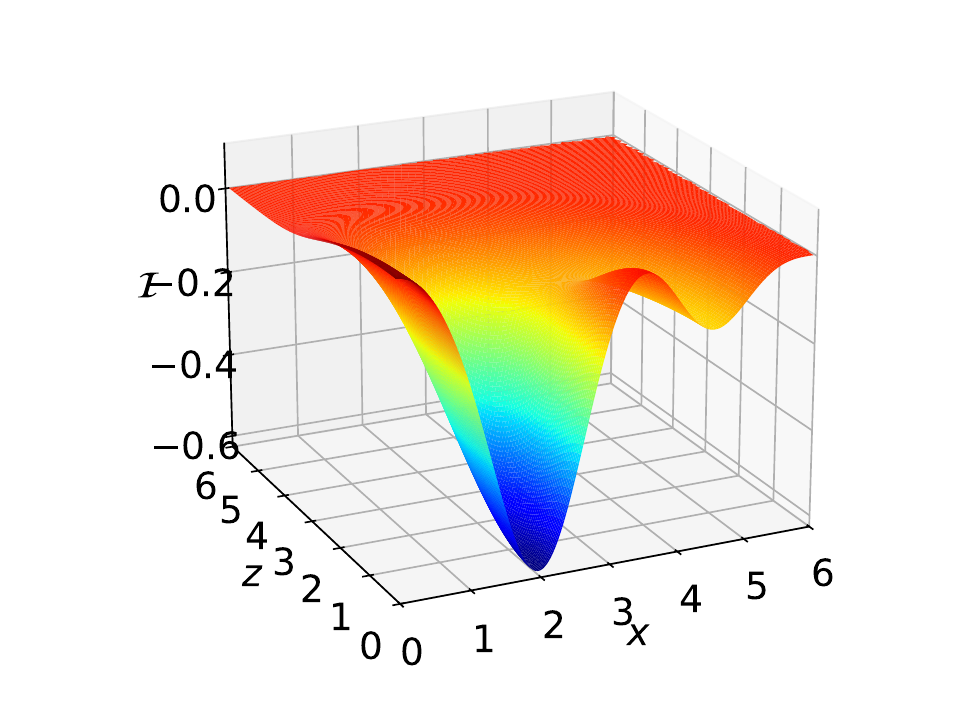}
\includegraphics[width = 0.31 \textwidth]{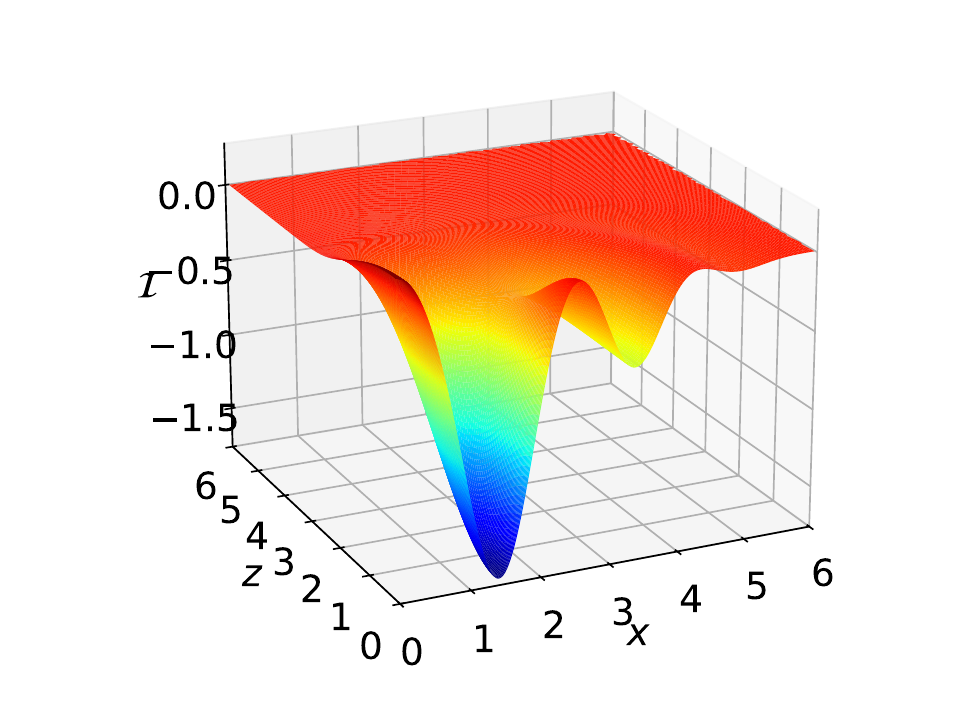}

\includegraphics[width = 0.31 \textwidth]{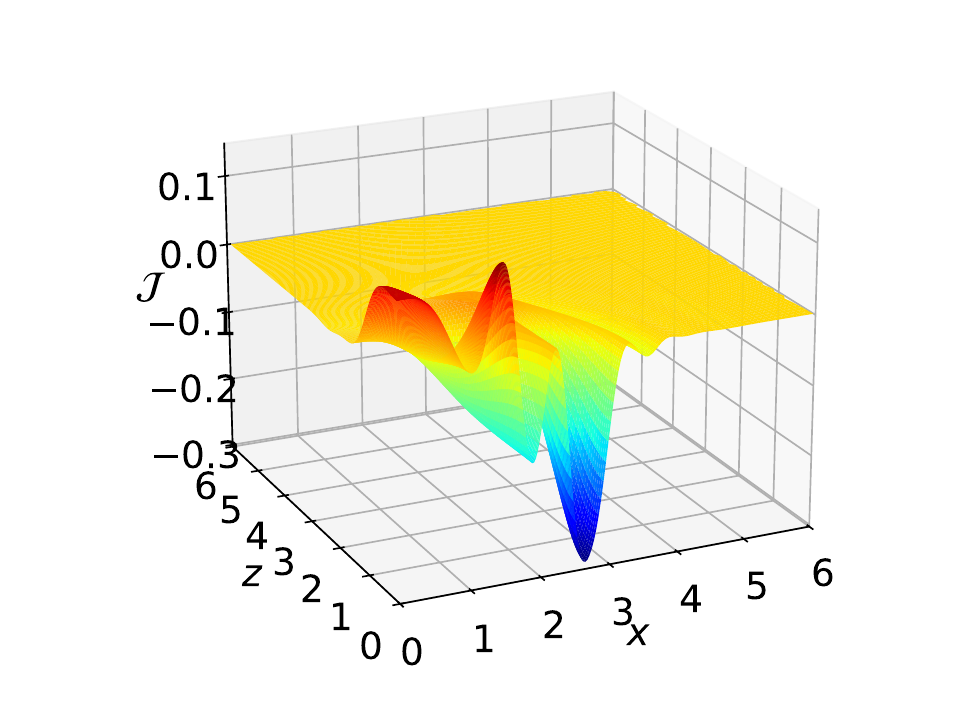}
\includegraphics[width = 0.31 \textwidth]{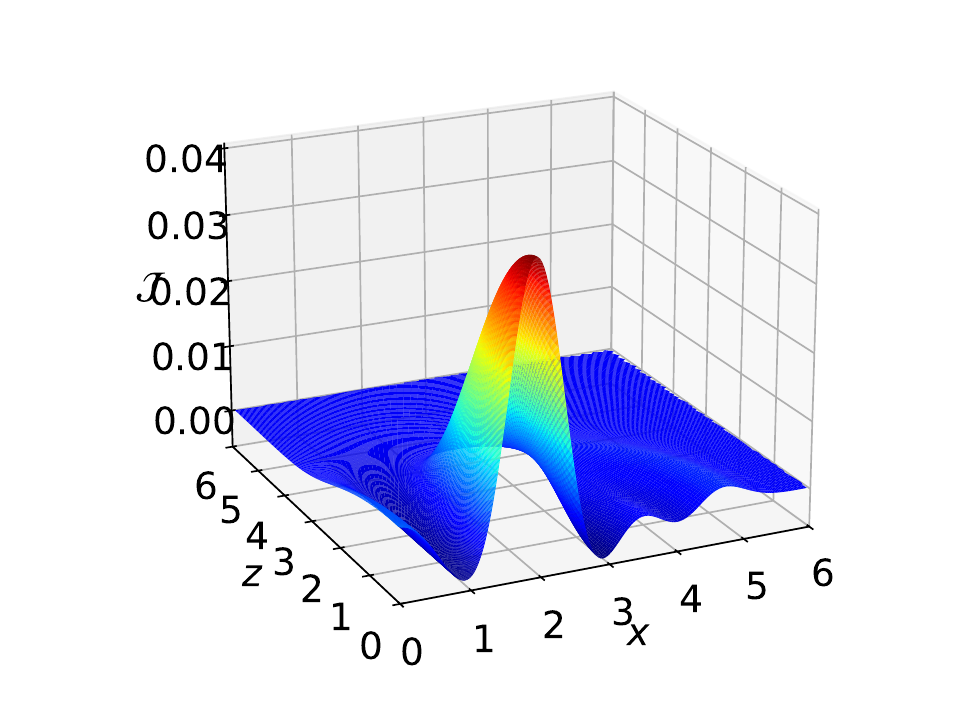}
\includegraphics[width = 0.31 \textwidth]{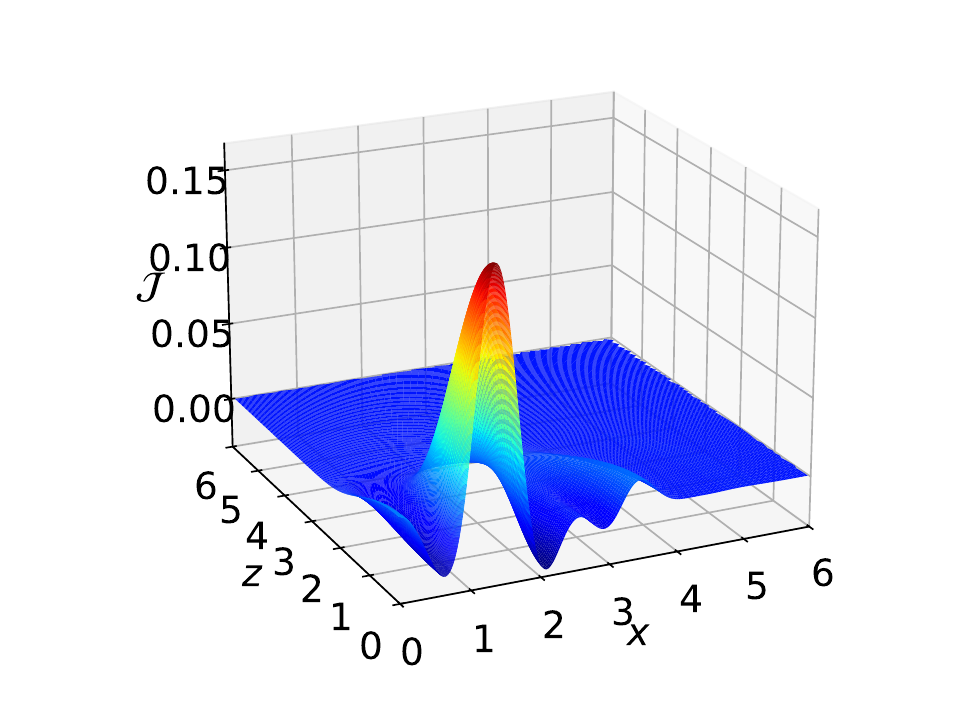}
\caption{Comparison of the curvature invariants ${\mathcal I}$ (top row) and ${\mathcal J}$ (bottom row) for the Teukolsky wave initial data of Section \ref{sec:teukolsky} (left column), and the two families $A$ and $B$ of Nakamura wave initial data of Section \ref{sec:nakamura} (middle and right panels).  Here $x = R \sin \theta$ and $z = R \cos\theta$, where $R$ is the proper distance from the center measured along a line of constant azimuthal angle $\theta$.  We show the data for the critical amplitudes ${\mathcal A}_*$ listed in Table \ref{tab:families} below.}
\label{fig:initial_data}
\end{figure*}

We show the curvature invariants ${\mathcal I}$ and ${\mathcal J}$ (see Section \ref{sec:diagnostics} below) for Teukolsky waves with the above choice of seed function, and with the amplitude ${\mathcal A}$ close to its critical value, in Fig.~\ref{fig:initial_data}.

%
\subsection{Nakamura waves}
\label{sec:nakamura}
%

In this paper we explore an alternative to the above construction.  We first observe that, rather than choosing the functions $F_+$ and $F_-$ in the Teukolsky formalism so the time derivative of their combination vanishes, one could also choose them so that the combination itself is zero, $F(0,r) = 0$, but the time derivative $\partial_t F(0,r)$ is not.  In this case the initial slice would have a flat spatial metric with non-zero extrinsic curvature.  This is the approach adopted in Nakamura \cite{Nak84} (see also \cite{ShiN95}), and we will therefore refer to these solutions as ``Nakamura" waves in order to distinguish them from those discussed in Section \ref{sec:teukolsky}.  

Like the Teukolsky waves of Section \ref{sec:teukolsky}, Nakamura waves are constructed from seed functions prescribing ingoing and outgoing waves.  Following \cite{Nak84} we will refer to these seed functions as $P_{\ell m}(t - r)$ and $Q_{\ell m}(t + r)$ for the even-parity waves considered here, and we will consider two specific choices in Sections \ref{sec:FamA} and \ref{sec:FamB} below.  We refer to Appendix \ref{sec:appendix} for a direct comparison of these seed functions with those used in the Teukolsky formalism.   From the functions $P_{\ell m}(t - r)$ and $Q_{\ell m}(t + r)$ we compute a new function $A_{\ell m}(r,t)$, which involves $\ell$ radial derivatives of the seed functions (see Eq.~(33) in \cite{Nak84}, hereafter (N33)).  We then compute three functions $G_{\ell m}$, $B_{\ell m}$, and $F_{\ell m}$, which involves additional radial derivatives of $A_{\ell m}(r,t)$ (see Eqs.~(N34) -- (N36)).  Finally, products of these functions with scalar, vector, and tensor spherical harmonics (expressed in terms of derivatives of the spherical harmonics $Y^{\ell m}$) provide analytical expressions for the extrinsic curvature $K_{ij}$ (see Eqs.~(N8) -- (N16)).  By construction, the latter is trace-free -- so that $A_{ij} = K_{ij}$ and $K = 0$ -- and satisfies $D^{\rm flat}_j A^{ij} = 0$, where $D_i^{\rm flat}$ is the flat covariant derivative.  We reiterate that these solutions can also be constructed using the Teukolsky approach discussed in Section \ref{sec:teukolsky}.

Given a solution for $A_{ij}$, the spatial metric can, in principle, be found from integration over time (see (N3) and (N37) -- (N40)).  This integration involves arbitrary integration constants (that may be functions of the spatial coordinates), and we may therefore choose the (linear) spatial metric to be flat, $\gamma^{\rm Nak}_{ij} = \eta_{ij}$.

It turns out that it is quite easy to ``dress up" these data so that they provide nonlinear solutions to the constraint equations.  We start by identifying $\bar \gamma_{ij} = \gamma^{\rm Nak}_{ij} = \eta_{ij}$, so that the spatial metric is conformally flat initially.   We next observe that the momentum constraint (\ref{eq:mom}) is linear in the extrinsic curvature and, for maximal slicing $K = 0$, decouples from the Hamiltonian constraint (\ref{eq:ham}).  As long as the spatial metric is conformally flat (so that the operator $\bar D_i$ in Eq.~(\ref{eq:mom}) is the flat covariant derivative), we therefore obtain exact solutions to the momentum constraint by identifying the Nakamura solutions for the extrinsic curvature with $\hat A_{ij}$.  As for the Teukolsky data, we then have to solve only the Hamiltonian constraint (\ref{eq:ham}) numerically.  Now, however, the Laplace operator $\bar D^2$ in (\ref{eq:ham}) reduces to the flat Laplace operator, which simplifies the calculation considerably (in addition to having $\bar R = 0$).   Moreover, the conformal connection functions vanish identically at the initial time, $\bar \Gamma^i = 0$, further simplifying the construction of the initial data.

In the following two subsections we provide closed-form solutions for $\hat A_{ij}^{\rm Nak}$ at the initial time $t = 0$ for two different choices of the seed functions, which we will refer to as families $A$ and $B$.  In Fig.~\ref{fig:initial_data} we compare the curvature invariants ${\mathcal I}$ and ${\mathcal J}$ for these two families with those for the Teukolsky wave of Section \ref{sec:teukolsky}.

%
\subsubsection{Family A}
\label{sec:FamA}
%

Explicit solutions for $\hat A_{ij}^{\rm Nak}$ are provided in \cite{Nak84,ShiN95} for quadrupolar waves ($\ell = 2$) with $m = 2$.  Here we focus on axisymmetric solutions and therefore construct new solutions for $m = 0$.  Adopting seed functions similar to those of \cite{Nak84,ShiN95},
\begin{subequations}
\begin{align}
P_{20} & = {\mathcal A} \left( \frac{4 \pi}{5} \right)^{1/2} \frac{\partial}{\partial t} 
\exp\left(- (r - t)^2 / 2 \right) \\
Q_{20} & = - {\mathcal A}\left( \frac{4 \pi}{5} \right)^{1/2} \frac{\partial}{\partial t} 
\exp\left(- (r + t)^2 / 2 \right), 
\end{align}
\end{subequations}
where ${\mathcal A}$ is the amplitude, we find the extrinsic curvature 
\begin{widetext}
\begin{subequations} \label{hatAij_famA}
\begin{align}
\hat A_{rr}^{\rm Nak} & = {\mathcal A} e^{-r^2/2} (1 - 3\cos^2 \theta )\\ 
\hat A_{r\theta}^{\rm Nak}  & = {\mathcal A} e^{-r^2/2} r  \, (3 - r^2) \sin\theta \cos\theta \\
\hat A_{r\varphi}^{\rm Nak} & = 0  \\
\hat A_{\theta\theta}^{\rm Nak} & = \frac{{\mathcal A} }{4} e^{-r^2 / 2} r^2 \, \Big( - 2 + 6 \cos^2\theta  - (6 - 8 r^2 + r^4) \sin^2\theta \Big) \\
\hat A_{\theta\varphi}^{\rm Nak} & = 0 \\
\hat A_{\varphi\varphi}^{\rm Nak} & = \frac{{\mathcal A} }{4} e^{-r^2 / 2} r^2 \sin^2 \theta \,\Big(-2 + 6 \cos^2\theta + (6 - 8 r^2 + r^4) \sin^2 \theta \Big)
\end{align}
\end{subequations}
at the initial time $t=0$. 

%
\subsubsection{Family B}
\label{sec:FamB}
%

We build a second family of quadrupolar waves from the functions (compare \cite{Nak84,ShiN95})
\begin{subequations}
\begin{align}
P_{20} & = {\mathcal A}  \left( \frac{4 \pi}{5} \right)^{1/2} \frac{\partial}{\partial t} \left\{ \left( (r - t)^2 + 1 \right)
\exp\left(- (r - t)^2 / 2 \right) \right\}\\
Q_{20} & = - {\mathcal A}  \left( \frac{4 \pi}{5} \right)^{1/2} \frac{\partial}{\partial t} \left\{ \left( (r + t)^2 - 1 \right)
\exp\left(- (r + t)^2 / 2 \right) \right\},
\end{align}
\end{subequations}
in which case the extrinsic curvature becomes
\begin{subequations} \label{hatAij_famB}
\begin{align}
\hat A_{rr}^{\rm Nak} & = {\mathcal A}  e^{-r^2/2} (5 - r^2) (1 - 3\cos^2 \theta )\\ 
\hat A_{r\theta}^{\rm Nak} & = {\mathcal A}  e^{-r^2/2} r  \, (15 - 10 r^2 + r^4) \sin\theta \cos\theta \\
\hat A_{r\varphi}^{\rm Nak} & = 0  \\
\hat A_{\theta\theta}^{\rm Nak} & = \frac{{\mathcal A} }{8} e^{-r^2 / 2} r^2 \, \Big( -80 + 128 r^2 - 34 r^4 + 2 r^6 + 2 (60 - 68 r^2 + 17 r^4 - r^6) \cos^2 \theta \Big) \\
\hat A_{\theta\varphi}^{\rm Nak} & = 0 \\
\hat A_{\varphi\varphi}^{\rm Nak} & = \frac{{\mathcal A} }{8} e^{-r^2 / 2} r^2 \sin^2 \theta \,\Big( 40 - 120 r^2 + 34 r^4 - 2 r^6 + 2 r^2 ( 56 - 17 r^2 + r^4) \cos^2 \theta \Big)
\end{align}
\end{subequations}
at the initial time $t = 0$.
\end{widetext}

%
\section{Numerics}
\label{sec:numerics}
%

%
\subsection{Evolution}
\label{sec:evolution}
%

We solve the Hamiltonian constraint (\ref{eq:ham}) and then evolve the initial data using a numerical relativity code that employs spherical polar coordinates (see, e.g., \cite{BauMCM13} for details).  Specifically, the code solves the BSSN formulation \cite{NakOK87,ShiN95,BauS98} of Einstein's equations and handles the coordinate singularities at the center $(r = 0)$ and on the axis ($\sin \theta = 0$) of the coordinate system using a reference-metric formulation (see \cite{Bro09,Gou12}) together with a rescaling of all tensorial quantities (see \cite{BauMCM13,RucEB18}).   The version used for the simulations presented here uses eighth-order spatial finite differencing together with a fourth-order Runge-Kutta time integration.   The current version also employs {\tt openMP}, which allowed us to adopt a higher radial grid resolution (see below) and to fine-tune the Nakamura waves of Sect.~\ref{sec:nakamura} to slightly better precision than for our simulations of Teukolsky waves in \cite{BauGH23}.   While the code does not require any symmetry assumptions, we impose both axisymmetry and equatorial symmetry for the simulations here, which is consistent with the initial data of Section \ref{sec:indata}.

\begin{figure}
    \centering
    \includegraphics[width=0.95\linewidth]{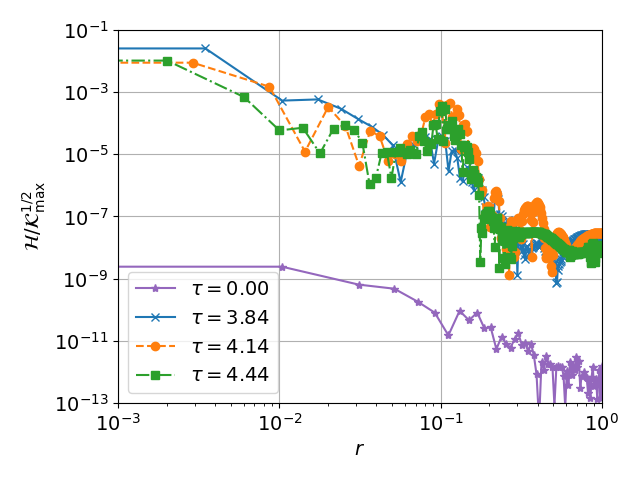}
    \caption{Violations ${\mathcal H}$ of the Hamiltonian constraint (\ref{eq:ham}) on the symmetry axis, for the near-critical Nakamura family $B$ evolution with ${\mathcal A}_* = 0.776404$.  In addition to the initial constraint violations we show results at the three times highlighted in Figs.~\ref{fig:rescaled_threshold} and \ref{fig:I_snapshots} below, i.e.~during the approximately self-similar part of the evolution.  We normalize ${\mathcal H}$ using the square root of maximum ${\mathcal K}_{\rm max}$ of the Kretschmann scalar (\ref{eq:Kretschmann}) at the current time.}
    \label{fig:Hamiltonian}
\end{figure}

The code has passed numerous numerical tests, including the convergence tests presented in, e.g., \cite{BauMCM13,BauGH19}.  In the context of vacuum gravitational waves, we have directly compared results from this code with those obtained with two completely independent codes (the {\tt prague} code used in \cite{LedK21} and the {\tt bamps} code used in \cite{FerRABH22}) and found excellent agreement (see \cite{Bauetal23,Adhetal26}).  Constraint violations remain small during the early part of the evolution, but grow during the latter part, in particular at the innermost grid point, which, in spherical polar coordinates, is most strongly affected by numerical noise.  We show snapshots of the violations of the Hamiltonian constraint in Fig.~\ref{fig:Hamiltonian}, both for the initial data at $\tau = 0$ and for three times during the late, approximately self-similar part of the evolution.

We carry out the integration using the shock-avoiding slicing condition of \cite{Alc03}, i.e.~we evolve the lapse function $\alpha$, starting with $\alpha = 1$ at the initial time $t=0$, using a Bona-Masso condition
\begin{equation} \label{eq:bonamasso}
(\partial_t - \beta^i \partial_i) \alpha = \alpha^2 f(\alpha) K
\end{equation}
(see \cite{BonMSS95}) with 
\begin{equation} \label{eq:shockavoiding}
f(\alpha) = 1 + \frac{1}{\alpha^2}.
\end{equation}
While this slicing condition does allow the lapse function to become negative (which is noticeable, for example, in Figs.~\ref{fig:threshold}, \ref{fig:rescaled_threshold}, and \ref{fig:rescaled_threshold_int} below), this does not appear to harm the evolution (see also \cite{BauH22,BeiC25}).  In (\ref{eq:bonamasso}), $\beta^i$ is the shift vector, which we choose to vanish in all simulations presented here.

Our code adopts a cell-centered non-uniform radial grid with $N_r$ gridpoints that is generated from a uniform grid in a variable $x \in (0,1)$ using the function
\begin{equation}
r = r_{\rm max} \frac{\sinh(s x)}{\sinh(s)},
\end{equation}
where $s > 0$ is a constant, and where $r_{\rm max}$ is the location of the outer boundary (see \cite{RucEB18}).  For $s \ll 1$ the grid approaches a uniform grid; otherwise the grid is nearly uniform close to the center at $r = 0$ but becomes approximately logarithmic for large $r$.  Our code also allows for regridding, by which we reduce the value of $r_{\rm max}$ from an initial value of $r_{\rm max}^{\rm init}$ to a final value of $r_{\rm max}^{\rm final}$ in $N_{\rm regrid} = 20$ steps, and continue the evolution with higher resolution inside $r_{\rm max}$.  We impose Robin-type boundary conditions on the outer boundaries, but terminate any evolution before the center is affected by the outer boundaries.   For the simulations presented here we used a uniform grid in the azimuthal angle $\theta$ with $N_\theta$ gridpoints.  

We previously evolved the family of Teukolsky initial data of Section \ref{sec:teukolsky} in \cite{BauGH23}, and we did not repeat those simulations.  Accordingly, our results for the Teukolsky waves were obtained with the parameters listed in \cite{BauGH23}, i.e.~$s = 4.0$, $N_r = 384$, $r_{\rm max}^{\rm init} = 32$, $r_{\rm max}^{\rm final} = 6.4$ and $N_\theta = 96$.  For the simulations of the Nakamura waves of Section \ref{sec:nakamura} we used the same values of $s = 4.0$ and $N_\theta = 96$, but chose $N_r = 640$, $r_{\rm max}^{\rm init} = 60$, and $r_{\rm max}^{\rm final} = 4$.

%
\subsection{Diagnostics}
\label{sec:diagnostics}
%

We diagnose the curvature of the spacetimes formed in the evolution of Teukolsky and Nakamura waves by evaluating the four invariants of the Weyl tensor $C_{abcd}$, which can be expressed as the real and imaginary parts of the two complex scalars $\I$ and $\J$ (see, e.g., \cite{SteKMHH03}).  We refer the reader to Sect.~II.C of \cite{BauGH23} on how the two scalars are computed in our code.  

For axisymmetric and twist-free spacetimes, the imaginary parts of both $\I$ and $\J$ vanish identically, and we will therefore report only real parts of $\I$ and $\J$ in the following.  Moreover, on the symmetry axis (where $\sin \theta = 0$), the two scalars satisfy $\I^3 = 27 \J^2$ identically.  We also note that the real part of $\I$ is related to the Kretschmann scalar ${\mathcal K}$ by
\begin{equation} \label{eq:Kretschmann}
{\mathcal K} \equiv {}^{(4)}R^{abcd}{}^{(4)}R_{abcd} = 16 \, \Re(\I), 
\end{equation}
where ${}^{(4)}R_{abcd}$ is the spacetime Riemann tensor.  As specific examples we show $\I$ and $\J$ for the initial data of the three families of Section \ref{sec:indata} in Fig.~\ref{fig:initial_data}.

In axisymmetry, and in the presence of a symmetry across the equator, the coordinate origin $r = 0$ represents the geodesic of a preferred observer.  It would therefore be natural to monitor the above curvature invariants at the coordinate origin.  However, in our code with spherical polar coordinates, these invariants (which are quadratic or cubic in components of the Riemann tensor) are also most strongly affected by numerical error at the origin, especially at late times when individual components of the Riemann tensor can become very large (compare Fig.~\ref{fig:Hamiltonian}).  We therefore consider two alternative measures that are less sensitive to numerical error at the origin, even though both of them introduce a slicing dependence.

First, we monitor the (local-in-time) maxima of the curvature invariants, by which we mean the maxima of the magnitude of $\I$ and $\J$ on a specific spatial slice $\Sigma$ corresponding to a coordinate time $t$,
\begin{equation} \label{eq:I_max}
\I_{\rm max}(t) \equiv \max_\Sigma \left(|\I(t,r,\theta)|\right)
\end{equation}
(and similar for $\J$ and ${\mathcal K}$).  We will distinguish the {\em global} maximum of the magnitude of $\I$,
\begin{equation}  \label{eq:I_MAX}
\I_{\rm MAX} \equiv \max_{t} \left(\I_{\rm max}(t)\right),
\end{equation}
by using capital letters in the subscript, and note that $|\I|_{\rm MAX}$ is no longer slicing-dependent.

While the maxima of $\I$ often do not occur at the center (see the discussion below), they typically occur close to the center, so that, at late times, $\I_{\rm max}$ can also be strongly affected by numerical error.  As a second alternative we therefore consider integrals of the curvature invariants over spatial slices, 
\begin{equation} \label{eq:I_int}
\I_{\rm int}(t) = \int_\Sigma |\I(t,r,\theta)| dV
\end{equation}
(where $dV$ is the (proper) volume element on $\Sigma$) which are less affected by numerical error at the innermost few grid points than the curvature itself at those grid points.  We will denote global maxima of the integrals (\ref{eq:I_int}) by
\begin{equation} \label{eq:I_int_MAX}
(\I_{\rm int})_{\rm MAX} \equiv \max_{t} \left(\I_{\rm int}(t)\right).
\end{equation}

We note that $\I$ has units of {\it length}$^{-4}$, so that $\I_{\rm int}$ has units of {\it length}$^{-1}$,  while $\J$ has units of {\it length}$^{-6}$.   

We focus on subcritical evolutions and therefore do not search for apparent horizons.

%
\section{Results}
\label{sec:results}
%

%
\subsection{Fine-tuning and scaling}
\label{sec:fine-tuning}
%

\begin{figure}[t]
    \centering
    \includegraphics[width=0.95\linewidth]{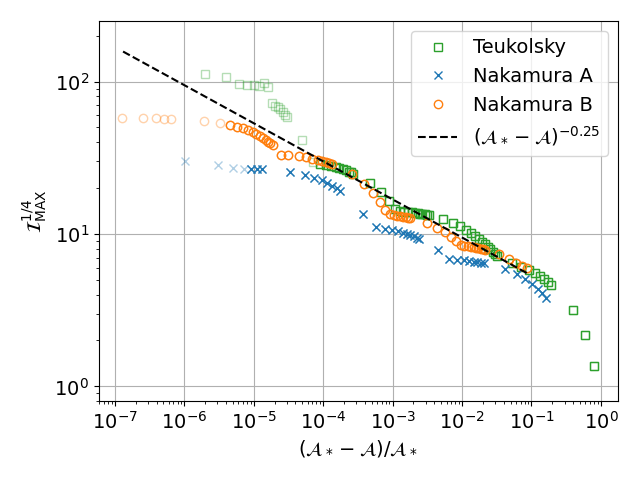}
    \caption{Scaling for our three families of initial data.  For each simulation performed we graph $\I_{\rm MAX}^{1/4}$ versus $({\mathcal A}_* - \mathcal A)/{\mathcal A}_*$, where ${\mathcal A}_*$ is our best estimate for the critical amplitude of the corresponding family (see Table \ref{tab:families}).  Here $\I_{\rm MAX}$ is the global maximum $\I_{\rm MAX}$ of the curvature invariant $\I$ attained during the simulation (see Eq.~(\ref{eq:I_MAX})).  Since $\I_{\rm MAX}^{1/4}$ has units of {\it length}$^{-1}$ it is expected to scale with $({\mathcal A}_* - {\mathcal A})^{-\gamma}$, where $\gamma$ is the critical exponent.  The dashed line represents this scaling with $\gamma = 0.25$.  Dots that are included as shaded only are results that may be affected by numerical error during the last part of the evolution.}
    \label{fig:I_MAX_scaling}
\end{figure}

\begin{figure}[t]
    \centering
    \includegraphics[width=0.95\linewidth]{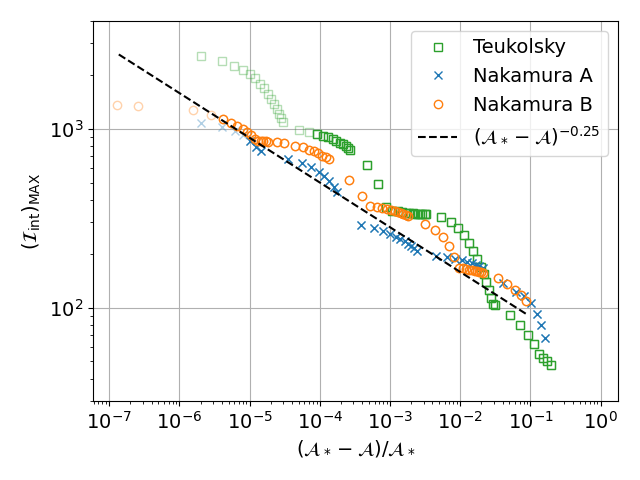}
    \caption{Same as Fig.~\ref{fig:I_MAX_scaling}, but for maxima of the integrals (\ref{eq:I_int_MAX}) of the scalar curvature $\I$.  }
    \label{fig:I_int_scaling}
\end{figure}

We start by fine-tuning the amplitudes ${\mathcal A}$ of the three initial data families of Sect.~\ref{sec:indata} to the threshold of black hole formation.  Knowing the critical amplitude ${\mathcal A}_*$ to $n$ digits, say, we determine the next digit by performing nine simulations for its nine possible values.  We distinguish subcritical from supercritical data both from the behavior of the curvature invariants and the lapse function at late times; for supercritical data, the former diverges while the latter drops to zero.  At late times, numerical error starts to dominate our diagnostics, which limits our ability to fine-tune to about six digits.  In the following we will report as the critical amplitude the largest amplitude ${\mathcal A}$ that resulted in a subcritical evolution (see, e.g., Table \ref{tab:families}).  

\begin{table}[t]
    \centering
    \begin{tabular}{l|c|c|c|c}
         Family & ${\mathcal A}_*$ & $\tau_*$ & $\Delta$ & $T_{\rm adjust}$ \\
         \hline
         Teukolsky & 0.00495634 & 4.35 & 0.67-0.73 & 1.1 \\
         Nakamura $A$ & 4.90186 & 7.4 & 0.5-0.85  & 0 \\
         Nakamura $B$ & 0.776404 & 5.3 & 0.52-0.88  & 0.3 \\
    \end{tabular}
    \caption{Parameters for our three families of initial data.  As an approximate value of the critical amplitude ${\mathcal A}_*$ we list for each family the largest subcritical amplitude; increasing the last digit by one then corresponded to the smallest supercritical amplitude in our simulations.  We also list estimates for $\tau_*$ computed from Eq.~(\ref{eq:tau_star}) for $\I_{\rm max}$ and adopted in Figs.~\ref{fig:rescaled_threshold} and \ref{fig:rescaled_threshold_int}, together with the range of values for $\Delta$ as obtained from (\ref{eq:Delta}).  In the last columns we list the values of $T_{\rm adjust}$ used in Figs.~\ref{fig:rescaled_threshold} and \ref{fig:rescaled_threshold_int}.}
    \label{tab:families}
\end{table}

Results for the Teukolsky waves of Sect.~\ref{sec:teukolsky} were discussed in \cite{BauGH23} already, where we found the strongest subcritical wave to have an initial amplitude of ${\mathcal A}_{*} \simeq 0.00495634$.  Here we follow a very similar procedure to locate the critical amplitudes for the Nakamura waves of Sect.~\ref{sec:nakamura} and find ${\mathcal A}_{*} \simeq 4.90186$ for family $A$ and ${\mathcal A}_{*} \simeq 0.776404$ for family $B$.  

We next examine our data for scaling, i.e.~we search for power laws of the form (\ref{eq:scaling_law}).  Specifically, we graph global quantities characterizing the spacetime evolution versus the distance of the wave amplitude ${\mathcal A}$ from its critical value ${\mathcal A}_*$ for all simulations performed.  Note that a quantity with units of {\it length}$^{n}$ is expected to scale with $({\mathcal A}_* - {\mathcal A})^{n \gamma}$ (compare (Eq.~\ref{eq:scaling_law})).  Since $\I_{\rm MAX}$ has units of {\it length}$^{-4}$, we plot in Fig.~\ref{fig:I_MAX_scaling} the fourth root of the global maximum $\I_{\rm MAX}$ attained during the evolution, which is expected to scale with $({\mathcal A}_* - {\mathcal A})^{- \gamma}$.  For comparison, we plot in Fig.~\ref{fig:I_int_scaling} global maxima $(\I_{\rm int})_{\rm MAX}$ of the integrals (\ref{eq:I_int}), which have units of {\it length}$^{-1}$ and therefore should also scale with $({\mathcal A}_* - {\mathcal A})^{- \gamma}$.  In both plots we include results for which the maximum attained curvature may be affected by the late-time numerical noise that we discussed in Sect.~\ref{sec:diagnostics} as shaded only.

While the details of the data for the three families differ from each other, all three families appear to display some approximate scaling.  It is difficult to determine an accurate critical exponent from the data in Figs.~\ref{fig:I_MAX_scaling} and \ref{fig:I_int_scaling}, but the scaling appears to be roughly consistent with $\gamma \simeq 0.25$, as indicated by the dashed line, for both $\I_{\rm MAX}$ and $(\I_{\rm int})_{\rm MAX}$.\footnote{The data for 
$(\I_{\rm int})_{\rm MAX}$ for the Teukolsky data in Fig.~\ref{fig:I_int_scaling} would also allow a slightly larger value of $\gamma$; compare the results reported in \cite{LedK21}.}  For a DSS critical solution we expect a periodic ``wiggle" to be superimposed on the simple power-law scaling (see \cite{HodP97,Gun97} as well as the discussion in Sect.~\ref{sec:intro}).  We indeed observe such wiggles in Figs.~\ref{fig:I_MAX_scaling} and \ref{fig:I_int_scaling}, but neither are the wiggles strictly periodic, nor are they the same between the different families.  This is an early indication that there is no unique critical solution, and that none of the different threshold solutions are strictly DSS.  

%
\subsection{Self-similarity and echoing}
\label{sec:echos}
%

\begin{figure}[t]
    \centering
    \includegraphics[width=0.99\linewidth]{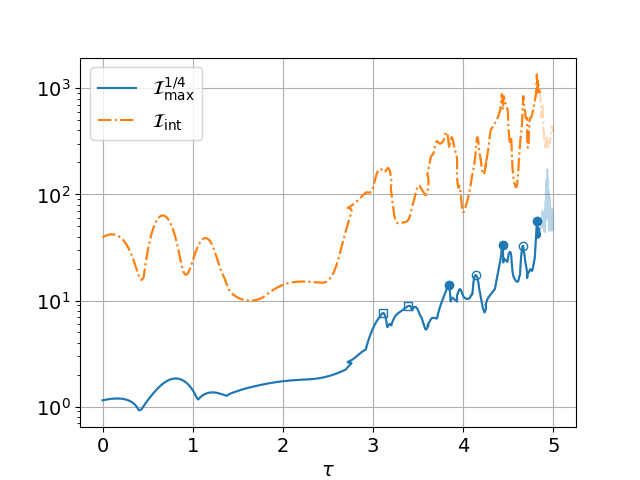}
    \caption{The threshold solution for the Nakamura family $B$ of Section \ref{sec:FamB}.  We show both $\I_{\rm max}$ (Eq.~\ref{eq:I_max}) and $\I_{\rm int}$ (Eq.~\ref{eq:I_int}) as a function of proper time $\tau$ as observed at the center.  The first part of the evolution, up to $\tau \simeq 1.5$ in code units, is dominated by the initial data, while  characteristics of the threshold solution emerge around $\tau \simeq 3$.  We highlight maxima (in time) of $\I_{\rm max}$ according to where (in space) that particular maximum is located.  The first two maxima are realized at the center, and are marked by open squares.  After that, a pattern emerges in which on-axis maxima on the axis (but away from the center, marked by filled circles) alternate with secondary maxima in the equatorial plane (also away from the center, marked by open circles).  The lightly shaded lines at late times are affected by numerical error, and are included for illustration only.}
    \label{fig:threshold}
\end{figure}

\begin{figure}[t]
    \centering
    \includegraphics[width=0.99\linewidth]{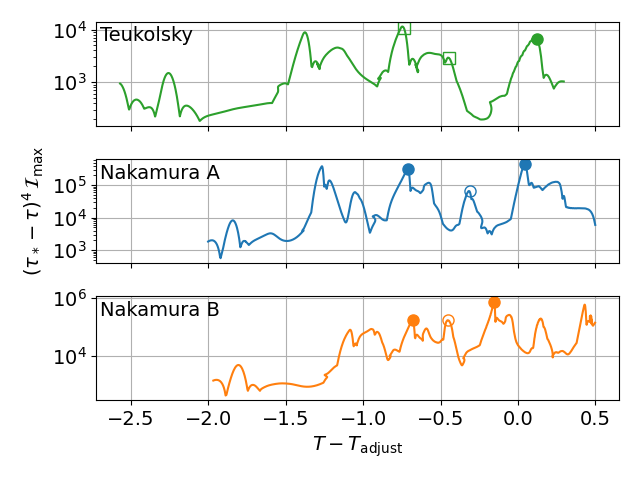}
    \caption{The maximum curvature $\I_{\rm max}$ (see \ref{eq:I_max}) a function of slow time $T$ (see \ref{eq:slow_time}) for the Teukolsky wave of Section \ref{sec:teukolsky} and the Nakamura waves of Section \ref{sec:nakamura}.  For each family we show results for the highest-amplitude subcritical solution that we simulated.  We multiply $\I_{\rm max}$ with $(\tau_* - \tau)^4$, so that, for a strictly DSS threshold solution, the graphs should become strictly periodic.  The three markers in each panel show the times for which we show profiles in Fig.~\ref{fig:I_snapshots}.}
    \label{fig:rescaled_threshold}
\end{figure}

\begin{figure}[t]
    \centering
    \includegraphics[width=0.99\linewidth]{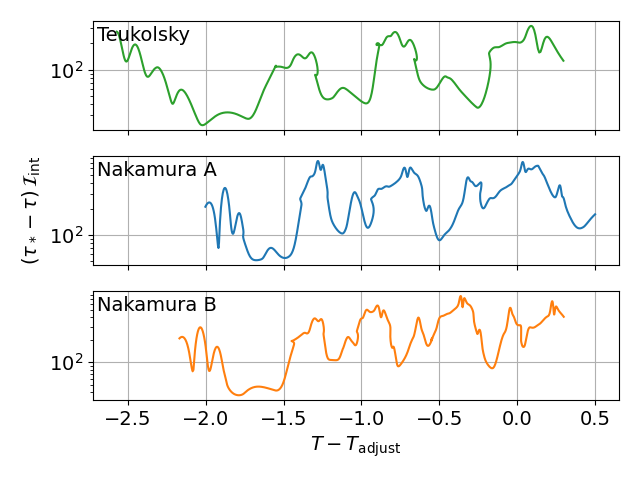}
    \caption{Same as Fig.~\ref{fig:rescaled_threshold}, but for the integrals $\I_{\rm int}$ (see \ref{eq:I_int}) rather than the maxima $\I_{\rm max}$.}
    \label{fig:rescaled_threshold_int}
\end{figure}

As an example of a threshold solution we show in Fig.~\ref{fig:threshold} the maximum curvature $\I_{\rm max}$ (see Eq.~\ref{eq:I_max}) and the integrated $\I_{\rm int}$ (see Eq.~\ref{eq:I_int}) as a function of proper time $\tau$ (as measured by an observer at $r=0$) for our highest-amplitude subcritical family $B$ Nakamura wave -- which is the family that we were able to fine-tune to highest precision.  The early part of the evolution, until $\tau \simeq 1.5$, is dominated by the initial data, while the characteristics of the threshold solution emerge around $\tau \simeq 3$.  Because our shock-avoiding slicing condition (\ref{eq:shockavoiding}) allows the lapse function to become negative, the proper time (as measured at the center) occasionally advances backwards for short periods of time.  This effect is visible in Fig.~\ref{fig:threshold} (as well as \ref{fig:rescaled_threshold} and \ref{fig:rescaled_threshold_int}), where the curves appear to be multi-valued for some times, or form small loops.

In Fig.~\ref{fig:threshold} we mark maxima in $\I_{\rm max}$ according to where (in space) these maxima occur.  The first two maxima, marked by open squares in Fig.~\ref{fig:threshold}, occur at the center, but later maxima do not.  Instead, the following maxima display an alternating pattern, in which maxima on the axis (marked by filled circles) are followed by maxima in the equatorial plane (marked by open circles) and vice versa.  In the following we will refer to the maxima on the axis, which generally are larger, as {\em on-axis maxima}, and those in the equatorial plane as {\em on-equator maxima}.  We will discuss this alternating pattern in more detail below; see, in particular, Figs.~\ref{fig:I_snapshots} and Figs.~\ref{fig:ax_photons} below.

As we discussed in Section \ref{sec:diagnostics}, our results for $\I_{\rm max}$ are affected by numerical error at late times, and we therefore consider our data reliable only up to a certain maximum time.  As an illustration of these effects we included some data beyond that maximum time as the lightly shaded lines in Fig.~\ref{fig:threshold}.  As anticipated in Section \ref{sec:diagnostics}, $\I_{\rm max}$ is more affected by these errors than the integrals $\I_{\rm int}$.

We next find approximate values for the proper time $\tau_*$ of the accumulation event as well as the echoing period $\Delta$.  In a CSS solution, i.e.~a self-similar solution that contracts continuously towards the accumulation event, a quantity $Q$ with units of {\it length}${}^n$ is expected to scale according to
\begin{equation} \label{eq:CSS_scaling}
Q \propto (\tau_* - \tau)^n
\end{equation} 
(where we take $\tau$ to be measured by an observer at the center $r=0$).  By contrast, a DSS solution is periodic with period $\Delta$ in what we refer to as {\em slow time}
\begin{equation} \label{eq:slow_time}
    T \equiv - \ln(\tau_* - \tau).
\end{equation}
Accordingly, the scaling (\ref{eq:CSS_scaling}) will apply only to values of $Q$ measured at times $T$ that differ by multiples of $\Delta$, which we refer to as {\em echos}.  For two subsequent echos we then have
\begin{equation}
\frac{Q_{i+1}}{Q_i} =
\left( \frac{\tau_* - \tau_{i+1}}{\tau_* - \tau_i} \right)^{n}
= \left( \frac{e^{-T_{i+1}}}{e^{-T_i}} \right)^{n} = e^{-n \Delta},
\end{equation}
where we have used $T_{i+1} = T_i + \Delta$ in the last step.

Knowing the values of $Q$ for two subsequent echoes, we can therefore find the echoing period $\Delta$ from
\begin{equation} \label{eq:Delta}
\Delta = - \frac{1}{n} \ln \frac{Q_{i+1}}{Q_i},
\end{equation}
and knowing the proper times $\tau$ of these echoes we obtain
\begin{equation} \label{eq:tau_star}
\tau_* = \frac{e^\Delta \tau_{i+1} - \tau_i}{e^\Delta - 1}
\end{equation}
(see also Sect.~III.A.2 of \cite{BauGH23}).

All of the above assumes, of course, that the threshold solution is perfectly DSS, fine-tuned to arbitrary precision, and unaffected by numerical error, none of which is the case for our data.  Even though our data may also be affected by gauge, we nevertheless compute approximate values of $\Delta$ and $\tau_*$ from the values and times of $\I_{\rm max}$ and $\I_{\rm int}$ at subsequent on-axis maxima identified in Fig.~\ref{fig:threshold} and similar graphs for the Teukolsky and Nakamura family $A$ waves. Applying (\ref{eq:Delta}) to three different pairs of $\I_{\rm max}$ for the Nakamura $B$ family, for example, results in values for $\Delta$ that range between about 0.52 and 0.88.  Similarly, Eq.~(\ref{eq:tau_star}) yields values for $\tau_*$ between about 4.7 and 5.4.  Discarding the earliest pair of data, for which the threshold solution has not yet reached the alternating pattern described above, we adopt $\tau_* \simeq 5.2$ as a compromise value, for example when computing the slow time $T$ in the following.  We apply similar arguments to the Teukolsky and Nakamura $A$ family, resulting in the values listed in Table \ref{tab:families} (see also the discussion in \cite{BauGH23}).

Adopting the values of $\tau_*$ as discussed above we can now graph characteristics of the threshold solution as a function of slow time $T$.  In particular, we graph in Fig.~\ref{fig:rescaled_threshold} the rescaled maximum curvature $(\tau_* - \tau)^4 \I_{\rm max}$ versus the slow time $T$, which, for a perfectly DSS solution, should result in a strictly period function.  Moreover, if there were a universal critical solution, all three families of initial data should converge to the same function at sufficiently late times and for sufficient fine-tuning.  While our results do not show evidence for either a universal or a strictly periodic threshold solution, the different threshold solutions do show some similarities.  In order to highlight these similarities we subtract from $T$ a value $T_{\rm adjust}$ chosen so that the later on-axis maxima occur at similar values of $T - T_{\rm adjust}$ (see Table \ref{tab:families} for specific values).  We then observe that all three families display similar patterns in terms of on-axis and on-equator maxima (even though, for the Teukolsky data, more of the early maxima occur at the center than for the Nakamura data), and that they all have similar values of the echoing period $\Delta$.  For comparison we show in Fig.~\ref{fig:rescaled_threshold_int} results for the integrated curvature $\I_{\rm int}$ rather than the maxima $\I_{\rm max}$.

%
\subsection{Spatial profiles}
\label{sec:spatial_profiles}
%

\begin{figure*}[t]
\centering
Teukolsky\hspace{1.5in}
Nakamura A\hspace{1.5in}
Nakamura B

\includegraphics[width = 0.31 \textwidth]{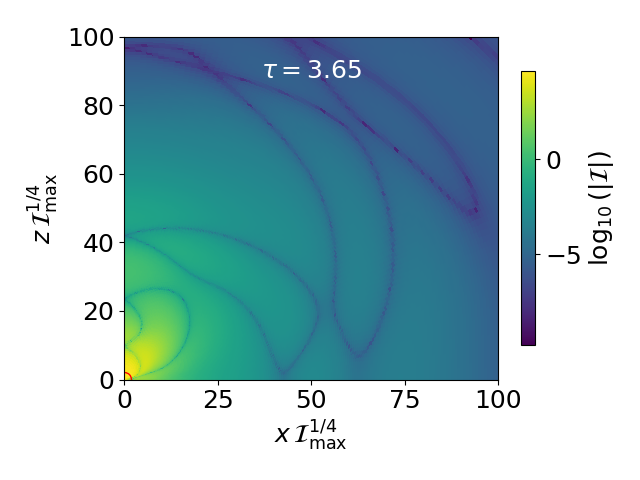}
\includegraphics[width = 0.31 \textwidth]{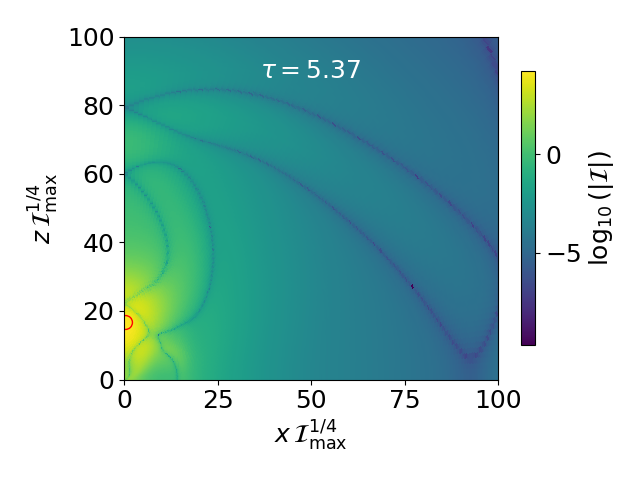}
\includegraphics[width = 0.31 \textwidth]{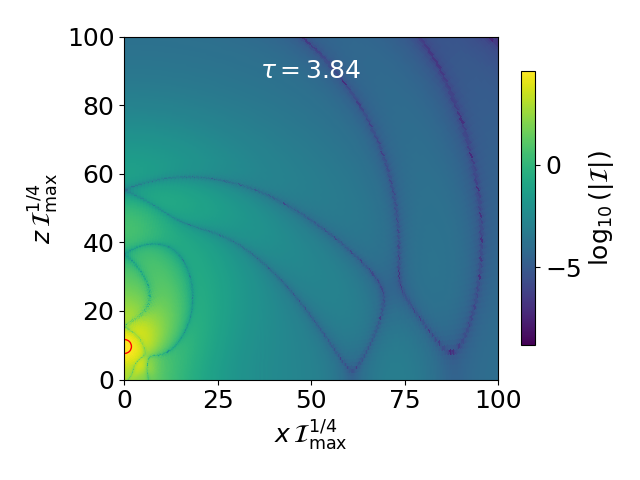}

\includegraphics[width = 0.31 \textwidth]{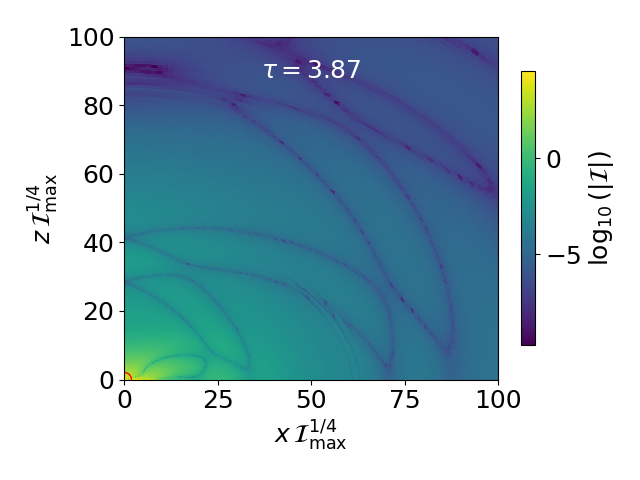}
\includegraphics[width = 0.31 \textwidth]{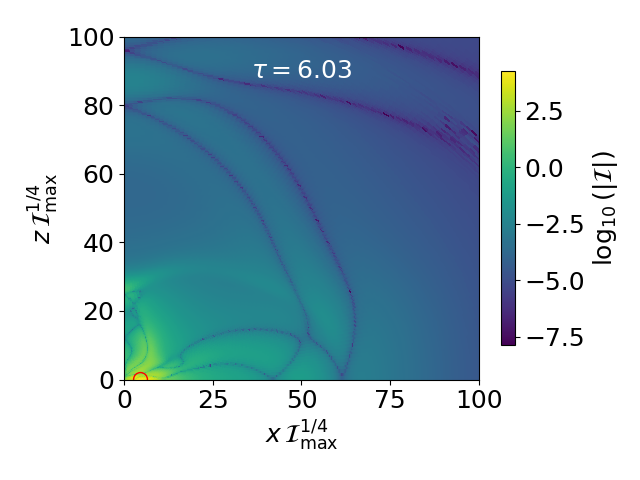}
\includegraphics[width = 0.31 \textwidth]{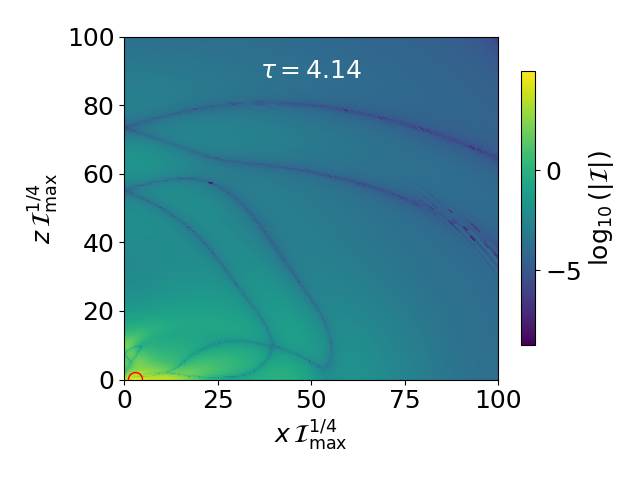}

\includegraphics[width = 0.31 \textwidth]{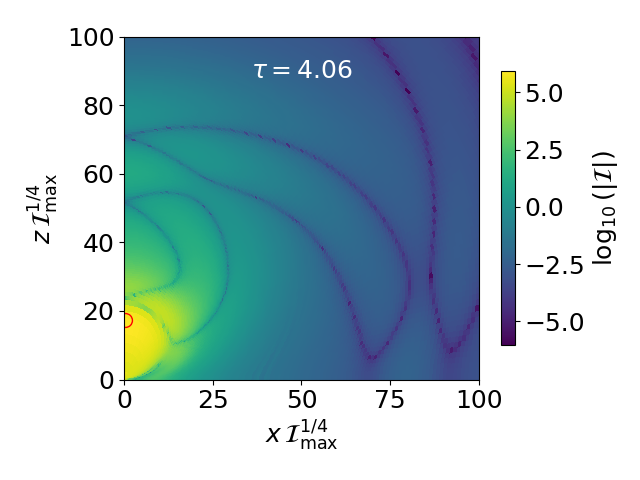}
\includegraphics[width = 0.31 \textwidth]{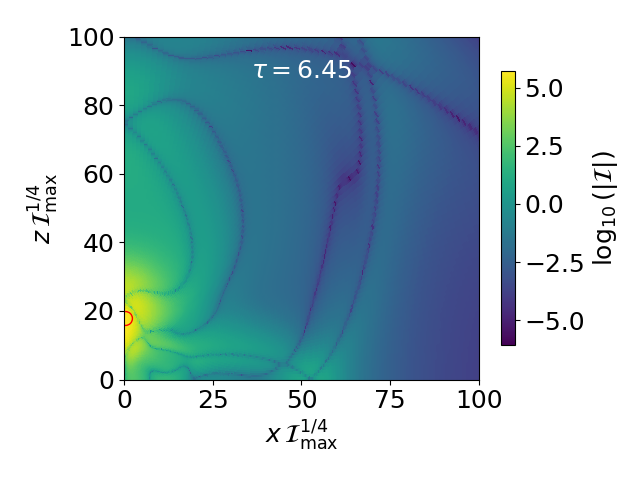}
\includegraphics[width = 0.31 \textwidth]{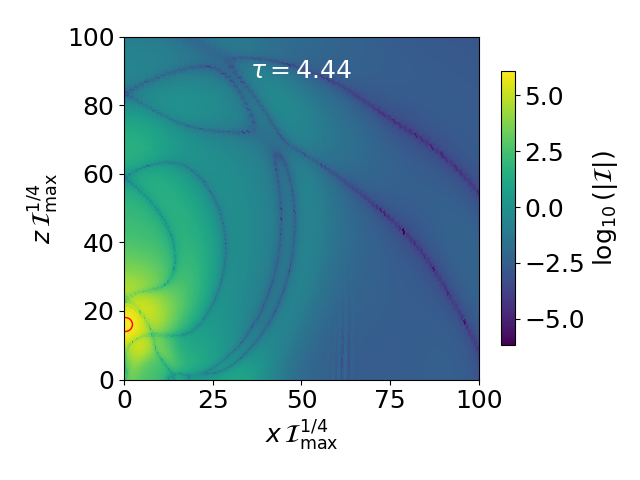}
\caption{Contour plots in the $xz$ plane of the magnitude of the curvature invariant ${\mathcal I}$, at times as close to those indicated by dots in Fig.~\ref{fig:rescaled_threshold} as possible.  The red circles mark the location of the maximum value $\I_{\rm max}$ at the respective times. The dark lines correspond to zero crossings of $\I$. }
\label{fig:I_snapshots}
\end{figure*}

In Fig.~\ref{fig:rescaled_threshold} we mark, for each family, three maxima of $\I_{\rm max}$ at similar stages of the evolution.  As in Fig.~\ref{fig:threshold}, we use different markers depending on the spatial location of the corresponding maximum: open squares at the center, filled circles on the axis, and open circles in the equatorial plane.\footnote{We note that the middle maximum for the Teukolsky wave indeed marks an on-equator maximum of $\I_{\rm max}$, even though it does not appear to be a maximum of $(\tau_* - \tau)^4 \,\I_{\rm max}$.}   In Fig.~\ref{fig:I_snapshots} we show spatial profiles of the curvature invariant $\I$ at these times (or as close to these times as data were available).  

The spatial coordinates $x = R \sin\theta$ and $z = R \cos\theta$ in Fig.~\ref{fig:I_snapshots} are computed using the proper distance from the center, as in Fig.~\ref{fig:initial_data}.  Here we multiply these coordinates with $\I_{\rm max}^{1/4}$ which, for a perfectly DSS solution, should result in periodic patterns.  We also mark the location of the maximum of $\I$ at each given time with the red circles.

As before we observe that, while the three families clearly show differences in their evolution, they also have some striking similarities.  Starting with the location of the maxima, the graphs in Fig.~\ref{fig:I_snapshots} confirm a pattern in which on-axis maxima (on the $z$-axis) alternate with on-equator maxima in the equatorial plane (on the $x$-axis) -- at least for the Nakamura families.  For the Teukolsky family, the first two maxima are still at the center (as earlier ones are for the Nakamura families as well, see Fig.~\ref{fig:threshold}), but the last one shown in Fig.~\ref{fig:I_snapshots} is also on the axis.

Away from the respective maxima, the profiles in Fig.~\ref{fig:I_snapshots} display a ``bean-like" pattern.  Note that the dark lines in the contour plots of Fig.~\ref{fig:I_snapshots} correspond to zero crossings of the curvature invariant $\I$.  These zero crossings form closed loops that, with some imagination, have a bean-like shape.  During the evolution, these patterns first appear alternatingly on the axis or in the equatorial plane, and then propagate outwards.  At early times, this behavior is qualitatively similar in all three families, even though the similarities are most noticeable between the Teukolsky and Nakamura $B$ families.  At later times, the zero crossings for the Teukolsky waves maintain the bean-like pattern -- hinting at an approximate DSS behavior even away from the maxima -- while the Nakamura waves develop more zero crossings and hence more complicated patterns.

%
\subsection{Spacetime profiles}
\label{sec:spacetime_profiles}
%

\begin{figure*}[t]
    \centering
    \includegraphics[width=0.49\linewidth]{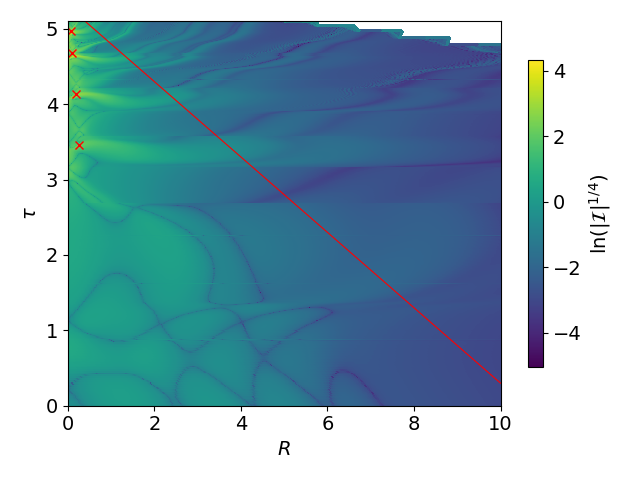}
    \includegraphics[width=0.49\linewidth]{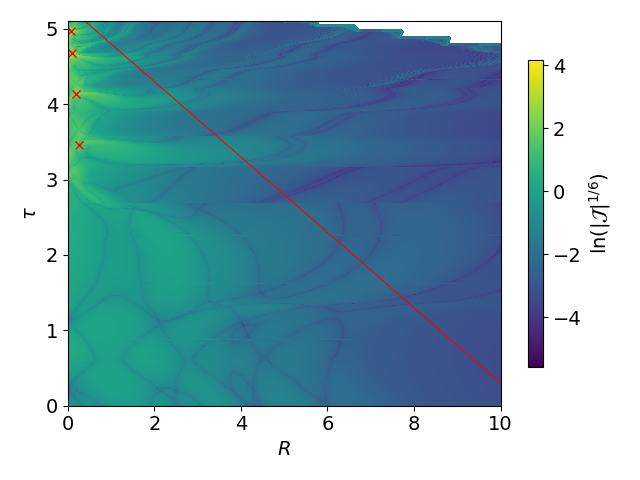}
    \caption{Spacetime profiles of the curvature invariants $\I$ (left panel) and $\J$ (right panel) for the near-critical Nakamura family $B$ in the equatorial plane, plotted as functions of the proper time $\tau$ (as observed at the center) and proper distance $R$ from the origin.  We include the red crosses to highlight some repeating patterns, and mark the same spacetime events in Figs.~\ref{fig:eq_profiles_rescaled} though \ref{fig:ax_photons} below.  The red lines mark the outer boundary $\xi_{\rm max} = 2$ of Fig.~\ref{fig:eq_profiles_rescaled}. (See the top two panels in Fig.~6 in \cite{BauGH23} for similar plots for the Teukolsky waves.)}
    \label{fig:eq_profiles}
\end{figure*}

\begin{figure*}[t]
    \centering
    \includegraphics[width=0.49\linewidth]{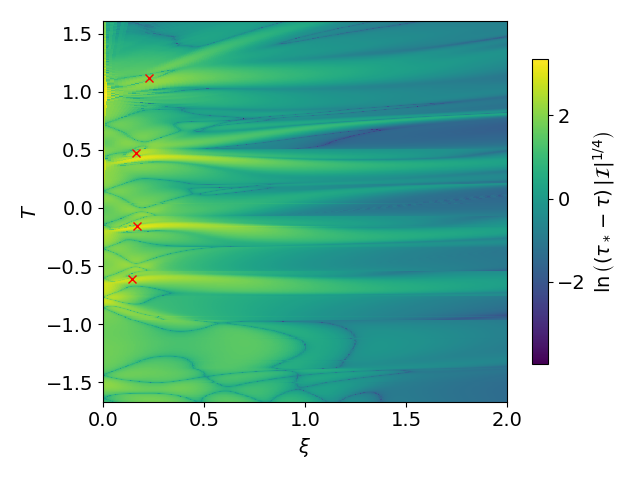}
    \includegraphics[width= 0.49\linewidth]{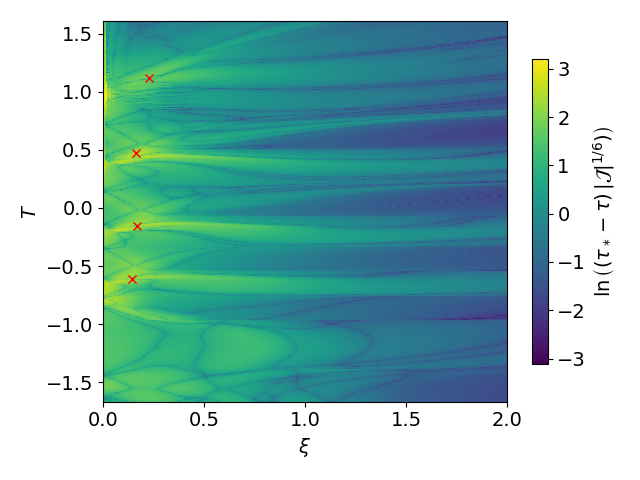}
    \caption{Same as Fig.~\ref{fig:eq_profiles}, except that we multiply $\mathcal{I}^{1/4}$ with $(\tau_* - \tau)$ and plot the result as functions of the slow time $T$ and the self-similar radius $\xi$.  (See the bottom two panels in Fig.~6 in \cite{BauGH23} for similar plots for the Teukolsky waves.)}
    \label{fig:eq_profiles_rescaled}
\end{figure*}

\begin{figure*}[t]
    \centering
    \includegraphics[width=0.49\linewidth]{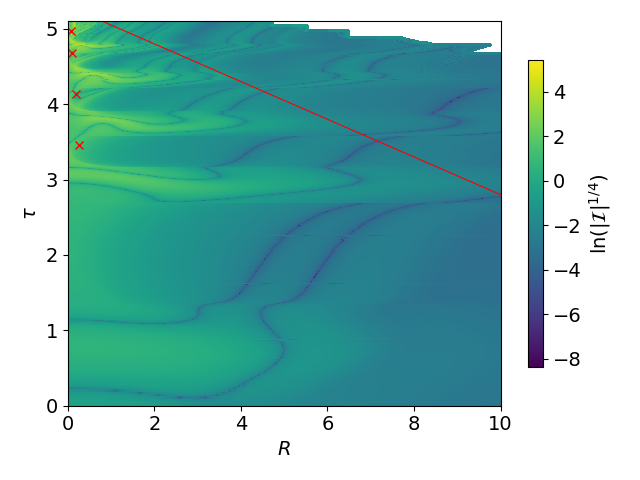}
    \includegraphics[width= 0.49\linewidth]{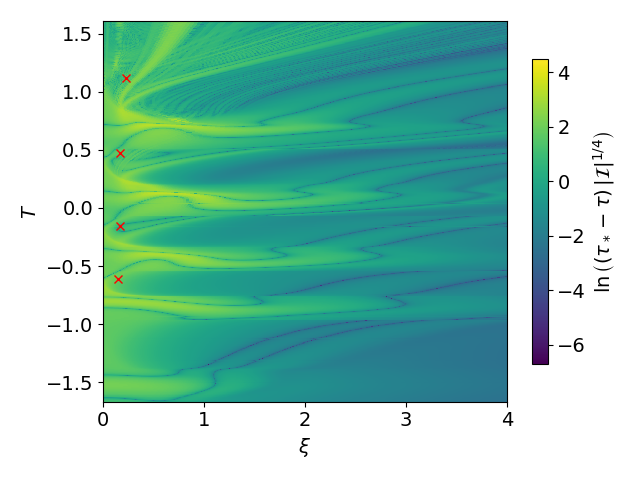}
    \caption{Same as Figs.~\ref{fig:eq_profiles} and \ref{fig:eq_profiles_rescaled}, but for results on the on the axis of symmetry rather than in the equatorial plane.  Since ${\mathcal I}$ and ${\mathcal J}$ do not contain independent information on the symmetry axis (see Section \ref{sec:diagnostics}) we show results for the former only.  The red crosses correspond to the same values of $\tau$ and $R$ (as well as $T$ and $\xi$) as those in Figs.~\ref{fig:eq_profiles} and \ref{fig:eq_profiles_rescaled}, but now appear between curvature maxima rather than on maxima.}
    \label{fig:axis_profiles}
\end{figure*}

We next turn to spacetime profiles, by which we mean graphs of the curvature invariants $\I$ and $\J$ as a function of the radius and time at fixed values of the azimuthal angle $\theta$.  We focus on the Nakamura family $B$ here, and refer to \cite{BauGH23} for similar plots for the Teukolsky family.

We start in Fig.~\ref{fig:eq_profiles} with profiles in the equatorial plane, i.e.~for $\theta = \pi/2$.  The early part of the evolution, until about $\tau \simeq 1.5$, shows both ingoing and outgoing zero crossings, which is not surprising given that the initial data describe a superposition of ingoing and outgoing waves.  At later times, after about $\tau \simeq 3$, a different pattern emerges, which we identify with properties of an approximately DSS threshold solution.  We include the red crosses (which were chosen as discussed below) to highlight some (approximately) repeating patterns. For the Teukolsky waves we found similar features (see Fig.~6 in \cite{BauGH23}), even though there the initial patterns resulting from the superposition of ingoing and outgoing waves persisted longer and were easier to identify -- possibly that is because we considered ``off-centered" Teukolsky waves, for which the initial curvature featured more wave cycles and a maximum further away from the center, than for the ``centered" Nakamura waves (compare Fig.~\ref{fig:initial_data}).

\begin{figure*}[t]
    \centering
    \includegraphics[width=0.49\linewidth]{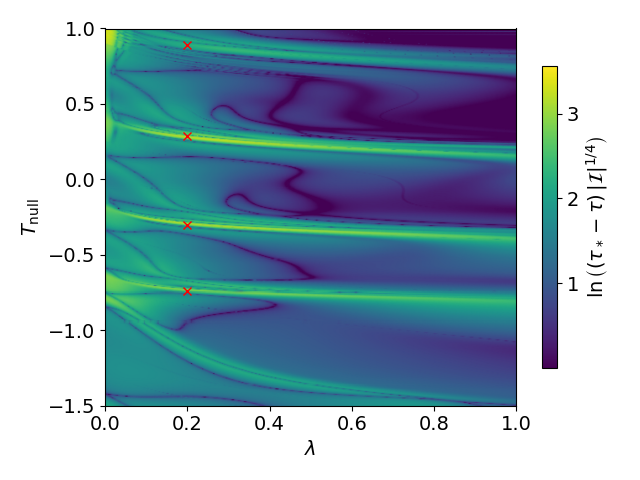}
    \includegraphics[width= 0.49\linewidth]{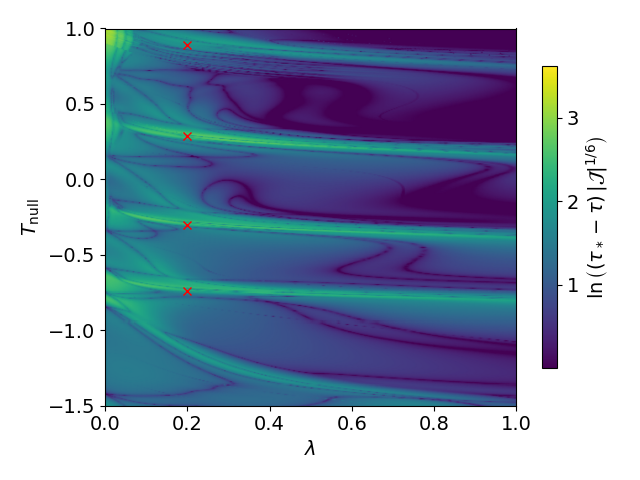}
    \caption{Spacetime profiles of the curvature invariants $\I$ (left panel) and $\J$ (right panel) in the equatorial plane, displayed in terms of outgoing null coordinates (see text for details).  The red crosses highlight repeating curvature maxima at $\lambda = 0.2$.  (Compare Fig.~8 in \cite{BauGH23}.) }
    \label{fig:eq_photons}
\end{figure*}

\begin{figure}[t]
    \centering
    \includegraphics[width=0.98\linewidth]{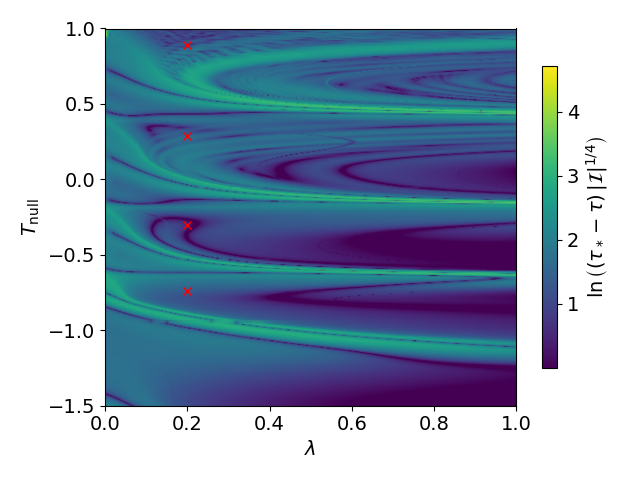}
    \caption{Spacetime profiles of the curvature invariant $\I$ on the axis in terms of outgoing null coordinates.  The red crosses mark the same values of $T_{\rm null}$ and $\lambda$ as in Fig.~\ref{fig:eq_photons}; note that here they appear {\em between} curvature maxima, highlighting the alternating pattern of maxima appearing on the axis and in the equatorial plane.  (Compare Fig.~7 in \cite{BauGH23}.) }
    \label{fig:ax_photons}
\end{figure}

In order to better analyze the self-similarity properties we show in Fig.~\ref{fig:eq_profiles_rescaled} the same data again, but as a function of slow time $T$ (see Eq.~(\ref{eq:slow_time})) and the self-similar radius
\begin{equation} \label{eq:xi}
\xi \equiv \frac{R}{\tau_* -\tau},
\end{equation}
and with $\mathcal{I}^{1/4}$ and $\mathcal{J}^{1/6}$ multiplied with $(\tau_* - \tau)$.  For a strictly DSS threshold solution -- and assuming that the slicing condition reflects the self-similarity of the solution -- the resulting plot should show a strictly periodic pattern.  Fig.~\ref{fig:eq_profiles_rescaled} does indeed show striking similarity between repeating patterns, indicating that our choice of coordinates reflects the approximate self-similarity to some degree. The red crosses included in Fig.~\ref{fig:eq_profiles} correspond to the same spacetime events as those marked in Fig.~\ref{fig:eq_profiles_rescaled}.   

In Fig.~\ref{fig:axis_profiles} we show profiles in on the axis of symmetry rather than in the equatorial plane.  Since $\mathcal{I}$ and $\mathcal{J}$ are not independent on the symmetry axis, we show profiles of the former only.  The red crosses mark the same values of $\tau$ and $R$ (and $T$ and $\xi$) as in Figs.~\ref{fig:eq_profiles} and \ref{fig:eq_profiles_rescaled}, but now appear between curvature maxima rather than on maxima.

As a third way of displaying the same data we graph in Fig.~\ref{fig:eq_photons} the curvature invariants as functions of null coordinates, which have a gauge-invariant meaning.  Specifically, we consider null geodesics that are emitted from the center.  We label these geodesics by the slow time $T_{\rm null}$ at which they are emitted, and an affine parameter $\lambda$ chosen so that $\lambda = 0$ and $(d\lambda / d\tau)_{T_{\rm null}} = dT / d\tau = (\tau_* - \tau)^{-1}$ at the center (so that $\lambda$ advances at the same rate as $T$ initially; see also \cite{BauGH19,BauGH23}).

Even though the plots in Fig.~\ref{fig:eq_photons} do not show an exact periodicity, it is easy to identify repeating patterns.  We used the red crosses to highlight maxima of the curvature at $\lambda = 0.2$; the red crosses in Figs.~\ref{fig:eq_profiles}, \ref{fig:eq_profiles_rescaled}, and \ref{fig:axis_profiles} mark the same spacetime locations.  

Finally, we show in Fig.~\ref{fig:ax_photons} the curvature $\I$ on the symmetry axis (as discussed in Sect.~\ref{sec:diagnostics}, we have $\I^3 = 27 \J^2$ on the axis, so that $\J$ does not contain independent information there). Similarly to the data in the equatorial plane, it is again possible to identify some repeating patterns.  We included the red crosses for the same values of $T_{\rm null}$ and $\lambda$ as in Fig.~\ref{fig:eq_photons}, and note that they now fall {\em between} maxima rather than on maxima.  This again demonstrates the alternating pattern of on-axis and on-equator maxima taken on the axis and in the equatorial plane.

%
\section{Discussion}
\label{sec:discussion}
%

Complementing numerous earlier studies (including \cite{AbrE93,AbrE94,Alcetal00,GarD01,Rin08,Sor11,Hiletal13,LedK21,FerRABH22,BauGH23,Bauetal23}) we report on numerical simulations of the critical collapse of vacuum gravitational waves.  Typically, these simulations have adopted one of two types of initial data, namely Brill waves \cite{Bri59}, which are based on a deformation of the conformally related spatial metric, or Teukolsky waves \cite{Teu82}, which are constructed from wave-like solutions to the linearized Einstein equations.  Contrary to critical phenomena observed in the gravitational collapse of spherically symmetric data, it appears that fine-tuning these different families of initial data to the onset of black hole formation does not result in a unique threshold solution, nor do any of the threshold solutions considered so far appear to be strictly DSS (see \cite{LedK21,FerRABH22,BauGH23,Bauetal23} as well as \cite{GunHMG25} for a review).  

In order to further explore the sensitivity of the threshold solution to the choice of initial data we consider in this paper a variation of the second type of initial data mentioned above.  Specifically, we again start with analytical solutions that describe linear gravitational waves, but, unlike in many previous approaches, we construct these data so that the spatial metric is flat at the initial time $t = 0$, while the wave content is described by the extrinsic curvature.  Even though these data can be constructed with the Teukolsky formalism of \cite{Teu82}, we follow the prescription of Nakamura \cite{Nak84} (see also \cite{ShiN95}) and therefore refer to these data as ``Nakamura" waves.

As we discuss in Sect.~\ref{sec:nakamura}, this approach leads to several simplifications.  Because the momentum constraint is linear, the extrinsic curvature satisfies this constraint exactly  --  even though it was adopted from a linear solution -- and being able to choose the initial time slice to be conformally flat makes it easier to solve the Hamiltonian constraint.

We consider two different families of axisymmetric Nakamura waves and fine-tune both to the onset of black hole formation to about 6 digits in the wave amplitudes.  While both threshold solutions show approximate scaling and an approximate DSS, the two threshold solutions do not agree with each other, nor do they agree with the threshold solutions previously found for Teukolsky or Brill waves. Our findings are therefore consistent with earlier conclusions, namely that, at least at the current level of fine-tuning, there is no evidence for a unique critical solution.  We also find that the departures from an exact DSS threshold solution appear to be larger for the two families of Nakamura waves than for the quadrupolar Teukolsky waves that we have previously considered in \cite{BauGH23}.  One might speculate whether this is related to the ``off-centered" or ``centered" nature of the respective initial data, but we have not explored that possibility here.

While previous authors have considered self-similar features of the curvature maxima that develop on the symmetry axis (see \cite{LedK21,FerRABH22}) we discuss here the appearance of on-axis and on-equator maxima that form alternatingly on the symmetry axis and in the equatorial plane.  The oscillation between these maxima appears to emit a curvature pattern from the central collapsing region whose zero crossings form the ``bean-like" shapes that we discussed in the context of Fig.~\ref{fig:I_snapshots}.  This observation suggests that any approximate self-similarity is not restricted to the curvature maxima themselves, but instead includes the entire central region of the collapsing spacetime.

\acknowledgments

This study was supported by the Oberwolfach Research Fellows program at the Mathematisches Forschungsinstitut Oberwolfach in 2022 and 2025; we greatly appreciate the institute's and its staff's hospitality and support during our stay.  This work was also supported in part by National Science Foundation (NSF) grant PHY-2308821 to Bowdoin College, as well as by FCT (Portugal) grant Numbers UID/00099/2025, UID/PRR/00099/2025 and 2023.12549.PEX. Numerical simulations were performed on the Bowdoin Computational Grid.

%

\appendix

%
\section{Comparison of Teukolsky and Nakamura formalisms for linear waves}
\label{sec:appendix}
%

In order to compare the seed functions used in the formalism of Teukolsky \cite{Teu82} and Rinne \cite{Rin09} with those used by Nakamura \cite{Nak84} we evaluate the functions $A = A(r,t)$ that govern the perturbations $\delta g_{rr}$ for even-parity vacuum linear perturbations of flat spacetime.\footnote{As in Sect.~\ref{sec:teukolsky} we adopt the notation and convention of \cite{Rin09} here, so that some expressions differ from those of \cite{Teu82} by constant factors.}  Identifying Eq.~(4) of \cite{Rin09} (hereafter (R4)) with (N7) we see that these functions are, in fact, identical,
\begin{equation}
\label{Acomparison}
A_{\rm{Rinne}} = A^h_{\rm{Nakamura}} 
\end{equation}
(here the superscript $h$ distinguishes the metric perturbation from the perturbation of the extrinsic curvature).

Restricting for simplicity to $\ell=2$, Appendix~A of \cite{Rin09}, interpreted in the light of the text between (R5) and (R6), yields a formula equivalent to
\begin{equation} \label{eq:RinneA}
A_{\rm Rinne} = -24\left({F_{,rr}\over r^3}-3{F_{,r}\over r^4}+3{F\over r^5}\right),
\end{equation}
where, to combine ingoing and outgoing solutions, we have defined
\begin{equation}
F(t,r) \equiv F_+(r+t)+F_-(r-t),
\end{equation}
and where the functions $F_+$ and $F_-$ are the same as in our discussion of Teukolsky waves in Sect.~\ref{sec:teukolsky}.
 
On the other hand, (N37) states
\begin{equation}
A^h_{\rm Nakamura} \equiv -2\int_{-\infty}^t A^K\,dt',
\end{equation}
where $A^K$ governs the perturbation $K_{rr}$.  Restricting (N33) to $\ell = 2$ then results in
\begin{equation}
A^K={{\cal F}_{,rr}\over r^3}-3{{\cal F}_{,r}\over r^4}+3{{\cal F}\over r^5}
\end{equation}
where we have abbreviated
\begin{equation}
{\cal F}(t,r) \equiv P(t-r)+Q(t+r),
\end{equation}
and where $P$ and $Q$ are as in \cite{Nak84} and in our discussion of Nakamura waves in Sect.~\ref{sec:nakamura} (except that we dropped the subscripts).  Note that the argument of $P$ is minus that of $F_-$.

Taking the time derivative of (\ref{Acomparison}) then yields
\begin{equation}
Q(x)=12 \, F_+'(x), \qquad P(x)=12\, F_-'(-x).
\end{equation}
We note that we have only compared the case $\ell=2$. For
general $\ell$, the expression (N33) is more concise than the
corresponding (R8a), and we have not checked explicitly whether they are
equivalent.

%

\end{document}